\documentclass{aa}
\pdfoutput=1

\usepackage{amsmath}
\usepackage[varg]{txfonts}
\usepackage{graphicx}
\usepackage{natbib}
\usepackage{xspace}
\usepackage{nicefrac}
\usepackage{multirow}
\usepackage{soul}

\defcitealias{De-Donder+2003}{DV03}

\newcommand{\kms}{\ensuremath{\,\rm{km}\,\rm{s}^{-1}}}
\newcommand{\Msun}{\ensuremath{M_{\odot}}\xspace}
\newcommand{\Lsun}{\ensuremath{L_{\odot}}}

\newcommand{\Myr}{\ensuremath{\,\mathrm{Myr}}\xspace}
\newcommand{\Myrs}{\ensuremath{\,\mathrm{Myr}}\xspace}

\newcommand{\Mmincc}{\ensuremath{\,M_{\min, \rm cc}}\xspace}

\authorrunning{E. Zapartas, S. E. de Mink,  R.G. Izzard,  S.-C. Yoon, C. Badenes et al.}
\titlerunning {Delay-time distribution of core-collapse supernovae}

\begin{document}

\title{ Delay-time distribution of core-collapse supernovae with late events resulting from binary interaction }

\author{E. Zapartas\inst{1} \and S.\,E. de Mink\inst{1} \and R.\,G. Izzard\inst{2} \and  S.-C. Yoon\inst{3} \and C. Badenes\inst{4} \and Y. G\"otberg\inst{1} \and  A. de Koter\inst{1,5} \and C.\,J. Neijssel\inst{1} \and M.~Renzo\inst{1} \and A. Schootemeijer\inst{6} \and T.\,S. Shrotriya\inst{6}}

\institute{Anton Pannekoek Institute for Astronomy, University of Amsterdam, Science Park 904, 1098 XH, Amsterdam, The Netherlands\\  
  \email{E.Zapartas@uva.nl, S.E.deMink@uva.nl}
  \and Institute of Astronomy, University of Cambridge, Madingley Road, Cambridge CB3 0HA, UK
  \and Astronomy Program, Department of Physics and Astronomy, Seoul National University, Seoul, 151-747, Korea
  \and Department of Physics and Astronomy \& Pittsburgh Particle Physics, Astrophysics, and Cosmology Center (PITT-PACC), University of Pittsburgh, Pittsburgh, PA 15260, USA
  \and Institute of Astronomy, KU Leuven, Celestijnenlaan 200 D, B-3001 Leuven, Belgium
  \and Argelander-Institut f{\"u}r Astronomie, Universit{\"a}t Bonn, Auf dem H{\"u}gel 71, 53121 Bonn, Germany
  }

   \date{Accepted for publication in Astronomy \& Astrophysics}
   
\abstract{
Most massive stars, the progenitors of core-collapse supernovae, are in close binary systems and may interact with their companion through mass transfer or merging. 
We undertake a population synthesis study to compute the delay-time distribution of core-collapse supernovae, that is, the supernova rate versus time following a starburst, taking into account binary interactions. We test the systematic robustness of our results by running various simulations to account for the uncertainties in our standard assumptions.

We find that a significant fraction, $15^{+9}_{-8}$\%, of core-collapse supernovae are `late', that is, they occur 50-200 Myrs after birth, when all massive single stars have already exploded. These late events originate predominantly from binary systems with at least one, or, in most cases, with both stars initially being of intermediate mass ($4-8\Msun$). The main evolutionary channels that contribute often involve either the merging of the initially more massive primary star with its companion or the engulfment of the remaining core of the primary by the expanding secondary that has accreted mass at an earlier evolutionary stage. Also, the total number of core-collapse supernovae increases by $14^{+15}_{-14}$\% because of binarity for the same initial stellar mass.

The high rate implies that we should have already observed such late core-collapse supernovae, but have not recognized them as such. We argue that $\phi$ Persei is a likely progenitor and that eccentric neutron star -- white dwarf systems are likely descendants. Late events can help explain the discrepancy in the delay-time distributions derived from supernova remnants in the Magellanic Clouds and extragalactic type Ia events, lowering the contribution of prompt Ia events. We discuss ways to test these predictions and speculate on the implications for supernova feedback in simulations of galaxy evolution.

}


\maketitle

\section{Introduction}\label{Section:Intro}
Core-collapse supernovae (ccSNe) are bright explosions that mark the end of the lives of massive stars  \citep[e.g.,][]{Heger+2003,Smartt2009a} and the birth of neutron stars or black holes \citep[e.g.,][]{Ertl+2016}.  They play a crucial role as sources of chemical enrichment \citep[e.g.,][]{Arnett1973,Woosley+2002} and feedback, driving the evolution of their host galaxies \citep[e.g.,][]{Hopkins+2014}.  Their extreme brightness also allows us to use them as probes of star-forming galaxies out to appreciable redshifts \citep[e.g.,][]{Strolger+2015}. 
These explosions are usually attributed to stars with birth masses larger than approximately 8\Msun  \citep{Heger+2003}, although the exact value depends on model assumptions concerning core overshooting, stellar-wind mass-loss, and metallicity \citep[e.g.,][]{Poelarends+2008,Jones+2013,Takahashi+2013,Doherty+2015}.

Observing campaigns of young massive stars in our galaxy and the Magellanic Clouds show that a very large fraction have one or more companions, forming a close binary system where severe interaction between the stars during their lives is unavoidable \citep[e.g.,][]{Kobulnicky+2007,Mason+2009,Sana+2012,Chini+2012}.  Such interaction can be the exchange of mass and angular momentum through Roche-lobe overflow, common envelope evolution, and merging of the two stars \citep{Wellstein+1999,de-Mink+2013,De-Marco+2016}. This interaction can drastically affect the further evolution of both stars and thus the properties of their possible supernovae. Pioneers in modeling the effects on (samples of) ccSNe include \citet{Podsiadlowski+1992, De-Donder+2003a, Yoon+2010, Eldridge+2008,  Eldridge+2013}.

Our understanding of the endpoints of massive stars is radically being transformed by the rise of (automated) transient surveys, which enable the efficient detection of ccSNe and other transients in large numbers. Examples are the Lick Observatory Supernova Search \citep[LOSS,][]{Filippenko+2001}, the Palomar Transient Factory \citep[PTF,][]{Rau+2009,Law+2009}, and its near-future upgrade, the Zwicky Transient Facility, the All-Sky Automated Survey for SuperNovae \citep[ASAS-SN,][]{Shappee+2014}, Pan-STARRS \citep{Kaiser+2002}, and eventually the Large Synoptic Survey Telescope \citep[LSST,][]{Ivezic+2008}.  	
The datasets provided by these facilities will be large, but may not necessarily provide very detailed information about individual events, since this typically requires more intensive follow up,  to obtain spectra, for example. The large potential of these datasets will be the statistical constraints that they can provide, allowing for new constraints on theoretical models for both common and rare events. Fully harvesting these datasets will require adaptations from the theory side and thus predictions of the statistical properties for large samples will be needed.  

Motivated by the technological and observational developments, as well as the insight into the large importance of binarity, we have started a systematic theoretical investigation aiming to quantify the impact of binarity on the statistical properties expected for large samples of ccSNe.  This paper is the first in a series in which we describe the motivation and setup of our simulations (\autoref{Section:Model}).  In two papers that were completed ahead of this one, the lead authors of this team demonstrated the early application of these new simulations against observations of two individual events.   

In  \citet{Van-Dyk+2016}, we compared these simulations with new deep Hubble Space Telescope observations of the site of the now faded stripped-envelope type Ic supernova SN1994I in search of a surviving companion star. While no companion was detected, the data provided new strong upper brightness limits, constraining the companion mass to less than 10\Msun. This result is consistent with the theoretical predictions of our simulations and allowed a subset of formation scenarios to be ruled out. In \citet{Margutti+2016}, we used these simulations to interpret the multi-wavelength observations of supernova SN2014C which over the timescale of a year underwent a complete metamorphosis from an ordinary H-poor type Ib supernova into a strongly interacting, H-rich supernova of type IIn. These simulations helped us to estimate the possibility that the surrounding hydrogen shell originated from a prior binary interaction \citep[as opposed to ejection resulting from instabilities during very late burning phases, e.g., ][]{Quataert+2016}. 

In this paper, we focus on the distribution of the expected delay time between formation of the progenitor star and its final explosion, extending the work of \citet{De-Donder+2003a}.  We investigate how binary interaction affects the delay-time distribution of ccSNe. A significant fraction of ccSNe are expected to be `late', that is, they occur with delay times longer than approximately 50 \Myr, which is the maximum delay time expected for single stars.  We show that these late events originate from progenitors in binary systems with most of them being of intermediate mass.  
We discuss these late ccSNe in \autoref{Section:Results} and describe the various evolutionary channels that produce them.

We further describe the outcome of an extensive study of the robustness of our results against variations in the model assumptions and we compare with earlier work  in \autoref{sec:robustness}.  In \autoref{sec:Discussion}, we discuss (possible) observational evidence. We argue that the well known binary $\phi$ Persei provides a direct progenitor system that is expected to result in a late ccSN and we discuss how the observed eccentric neutron star -- white dwarf systems may well provide the direct remnants.  We then compare our results directly with the inferred delay time measured from supernova remnants in the Magellanic Clouds, showing that they are consistent with our predictions.  We finish with a brief discussion on possible implications for feedback in star-forming regions in galaxies by showing the differences with the widely used single star predictions by the STARBURST99 simulations \citep{Leitherer+1999}. We end with a summary of our findings in \autoref{sec:summary}.

\section{Method}\label{Section:Model}

We use a binary population synthesis code, {\tt binary\_c} (version 2.0, SVN revision 4105), developed by \citet{Izzard+2004,Izzard+2006,Izzard+2009} with updates described in \citet{de-Mink+2013} and \citet{Schneider+2015}. The code employs rapid algorithms by \citet{Tout+1997} and \citet{Hurley+2000, Hurley+2002} based on analytical fits to the detailed non-rotating single stellar models computed by \citet{Pols+1998}. 

The code enables us to efficiently simulate the evolution of single stars and binary systems from the zero-age main sequence until they leave behind compact remnants. This allows us to make predictions for an entire population of massive stars by spanning the extensive parameter space of initial properties that determine their evolution. It also allows us to explore the robustness of our results against variations in our assumptions.

In \autoref{subsec:initial_conditions}, we discuss the initial conditions and in \autoref{subsec:physical_assumptions}, we discuss the physical assumptions. From now on when we mention ``standard models'' or ``standard simulations'', we refer to the simulations where we followed our main assumptions in all the key parameters that we discuss below. There are two standard models with one simulating only single stars and the other including binaries (discussed also below in the paragraph for binary fraction).  A summary of the key parameters, their values for our standard assumptions and the model variations that we consider is provided in \autoref{table:parameters_standard}.

  \subsection{Initial conditions}
  \label{subsec:initial_conditions}
   
\paragraph{Initial distributions}-- We assume that the distribution of the initial mass, $M_1$, of primary stars (the initially most massive star in a binary system) and of single stars follows a  \citet{Kroupa2001} initial mass function (IMF),
  
   \begin{equation}\label{eq:IMF}
 \frac{\mathrm{d}N}{\mathrm{d}M_1} \propto M_{1}^{\alpha'} ,
  \end{equation}
  where,
    \begin{equation}
  \alpha'= \left  \{ 
  \begin{array}{r}
  -0.3 \\
    -1.3 \\ 
      -2.3 \\
   \alpha 
 \end{array} 
\quad   \quad  
  \begin{array}{rcl}
 \ 0.01< &M_1/\Msun&<0.08 \mathrm{, }\\
 \ 0.08< &M_1/\Msun&<0.5   \mathrm{, }\\
 \ 0.5< &M_1/\Msun&<1  \mathrm{,}\\
 \ 1< &M_1/\Msun&<100.
 \end{array} \right.
\label{eq:imf}
 \end{equation}
\noindent
In our standard models, we adopt $\alpha = -2.3$. When assessing the uncertainties, we consider variations in which $\alpha = -1.6$ and $-3.0$ following the uncertainty given in \citet{Kroupa2001} as well as one model in which  $\alpha = -2.7$ \citep[e.g.,][]{Kroupa+1993}.
 
For the {initial mass ratio} $q \equiv M_2/M_1$, where $M_2$ is the initial mass of the secondary star, we take
   \begin{equation}
 \frac{\mathrm{d}N}{\mathrm{d}q} \propto q^{\kappa}.
 \label{eq:iqf}
  \end{equation}
We adopt a uniform distribution in our standard simulation, for example, $\kappa = 0$ for $q \in [0.1, 1]$, consistent with \citet{Kiminki+2012} and  \citet{Sana+2012}. We also consider the variations  $\kappa = -1$ and $1$. 

For the {initial orbital period} distribution, we assume
  \begin{equation}
    \frac{\mathrm{d}N}{\mathrm{d}\log_{10}P} \propto \left( \log_{10} P \right)^{\pi}. 
  \end{equation}
We adopt $\pi = 0$, also known as \citeauthor{Opik1924}'s law (\citeyear{Opik1924}), for systems with primary masses up to 15\Msun  \citep{Kobulnicky+2014,Moe+2015a}. To account for the strong preference of more massive stars to reside in short period systems, we adopt $\pi = -0.55$  when $M_1 > 15\Msun$ as found by \citet{Sana+2012}. The range of initial periods we consider is $\log_{10} (P/\mathrm{day}) \in [0.15, 3.5]$ as given by \citet{Sana+2012}. When assessing the uncertainties, we consider $\pi = -1$ and $1$ over the full mass range.

For the initial spin period of the stars, we follow \citet{Hurley+2000}.  Although this does not account for the full distribution \citep[e.g.,][]{Huang+2010,Dufton+2013,Ramirez-Agudelo+2013,Ramirez-Agudelo+2015}, this is sufficient for investigating the role of binarity as the impact of the adopted birth spin is negligible compared to the angular momentum the star later receives as a result of interaction by tides and mass transfer \citep{de-Mink+2013}.  

Although we account for the effects of eccentricity, we chose to adopt circular orbits at birth to limit the number of dimensions that our grid of models spans.  This is justified as most systems circularize shortly before the onset of mass transfer by Roche-lobe overflow as a result of tides \citep{Portegies-Zwart+1996,Hurley+2002}.  However, with this approach, we do not account for systems that are too wide to strongly interact when circular, but where eccentricity implies periastron separations small enough to trigger Roche-lobe overflow.  We may therefore slightly underestimate the impact of binary interaction. See, for example, the interacting systems arising from binaries with initial orbital periods well in excess of $10^{3.5}$ days depicted in Fig.~2 of \citet{de-Mink+2015}. We explore the uncertainties arising from this assumption indirectly  when we vary the initial orbital period distribution and the total binary fraction. 

\paragraph{Binary fraction}-- In our two standard models we either simulate only single stars or we adopt a binary fraction of $f_{\rm bin} = 0.7$. Here, we define a binary as a system with initial mass ratio $q \in [0.1,1]$ and initial period  $\log_{10} (P/\mathrm{day}) \in [0.15, 3.5]$ based on  \citet{Sana+2012} and consistent with the ranges adopted above. We consider variations of  $f_{\rm bin} = 0.3$ and 1.0. 

The binary fraction for intermediate-mass stars is less well constrained.  We therefore consider a model variation where we adopt a binary fraction that decreases with mass based on \citet{Moe+2013}. These authors provide the inferred fraction of systems with a companion in very close orbit, $P= 2-10$ days, and $q>0.1$. They find that this fraction drops from 0.22 for early B-type to 0.16 for late B-type stars. Information on systems with orbits longer than 10 days are not available from this study. These results may either indicate that the binary fraction decreases towards late spectral types or that there is simply a preference for systems with orbital periods longer than 10 days in these stars.  We use these results to construct a mass-dependent binary fraction assuming  that the binary companions of B-type stars still follow an \citep{Opik1924} law over the full period range $0.15 < \log_{10} P < 3.5$.  We construct a mass-dependent binary fraction  $f_{\rm bin}(M)$ referring to the binary fraction for the full period range,  such that the binary fraction for periods $P= 2-10$ days are as in \citet{Moe+2013}. This results in 

    \begin{equation}
  f_{\rm bin} (M_1)= \left  \{ 
  \begin{array}{l}
  0.44 \\
    0.61 \\ 
      0.7 \\
   \end{array} \right.
\quad \quad
  \begin{array}{rcl}
          & M_1 /\Msun  &  <6 \\
  6 \le & M_1 /\Msun  & < 15 \\
 15 \le & M_1 /\Msun &   \\
 \end{array}
 \quad
  \begin{array}{lll}
\sim \text{late B,} \\
\sim \text{early B,} \\
\sim \text{O, }\\
 \end{array}  \label{eq:fbin_mass}
 \end{equation}
which we adopt as one of the model variations.  
We want to stress the importance of further observational campaigns aimed to constrain the initial binary distributions and the binary fraction for the full $M_1$, $q$ and period range \citep[e.g.,][]{Moe+2016}.

\paragraph{Normalization}-- 
When quoting absolute rates, we express our results normalized by the total mass formed in stars in units of $10^6 \Msun$. For this, we integrate over the full range of the IMF as specified in \autoref{eq:imf}. For stars with masses above $M_{{\rm low}}$, we account for the mass contained in the companion star as specified in \autoref{eq:iqf}.   Effectively, we assume that low-mass stars, with $M_{1} < M_{{\rm low}}$, do not have companions massive enough to significantly contribute to the mass of the stellar population. In our standard assumptions we adopt  $M_{{\rm low}} =2\Msun$ and we vary this parameter to 1 and 3\Msun to check that this choice does not have a large influence on our results.

    \paragraph{Metallicity}--
    We assume solar metallicity in our standard population, adopting a mass fraction of elements heavier than helium of $Z=0.014$ \citep{Asplund+2009} because present day transient surveys focus on larger galaxies with metallicities that are comparable to solar.  We also consider low metallicities of $Z = 2\times10^{-4}$ relevant for metal-poor progenitors of globular clusters and populations formed at high redshift. We additionally test  $Z = 0.004$ and $Z = 0.008$, relevant to nearby dwarf galaxies similar to the Small and Large Magellanic Clouds.  We further provide results for the former canonical solar abundance of $Z= 0.02$ \citep[e.g.,][]{Grevesse+1996} for comparison with earlier studies and one super solar metallicity, $Z=0.03$, relevant to the central regions of large galaxies.   
 
 \begin{figure}[htp]\center
\includegraphics[width=0.5\textwidth]{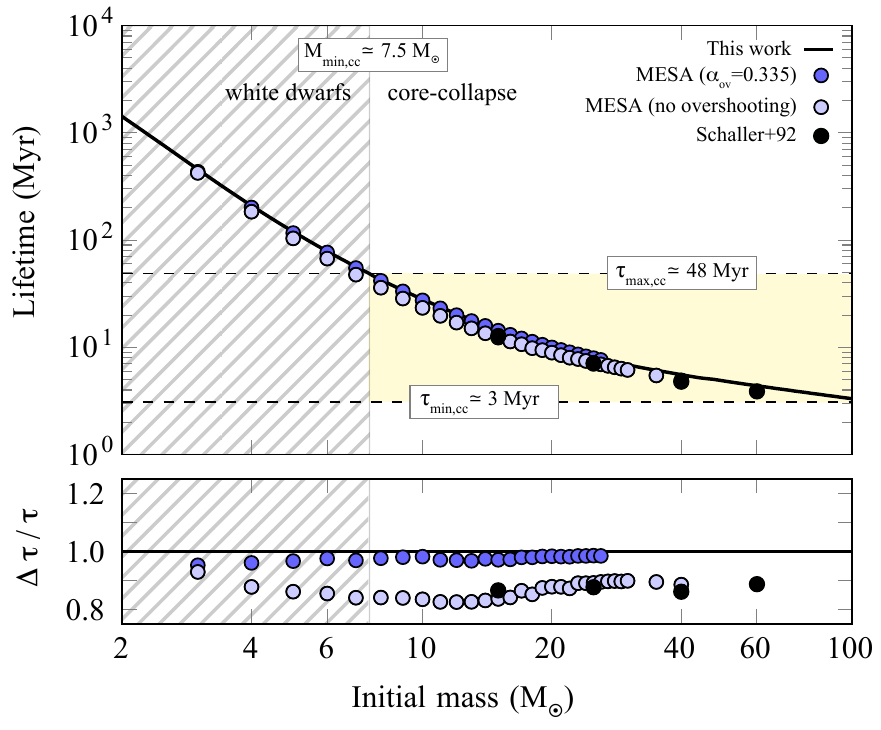}
\caption{ The lifetime $\tau$ (until the white dwarf phase or core collapse) as a function of initial mass for single stars adopted in this work (black line) is compared with predictions that we obtained using the MESA stellar evolutionary code \citep{Paxton+2011} with and without overshooting (dark and lightblue dots) and with Geneva models of \citet[][black dots]{Schaller+1992}.  Single stars with masses less than $M_{\min, \rm cc}$  end their lives as white dwarfs instead of ccSNe (hashed region).  ccSNe are expected between  $\tau _{\min,\mathrm{cc}}$ and $\tau _{\max,\mathrm{cc}}$ (yellow shaded region), which refer to the lifetimes of the most and least massive star to undergo core collapse.  
The bottom panel shows the relative difference in lifetimes with respect to the lifetimes used in our work (\autoref{subsec:physical_assumptions}).
\label{fig:lifetime}}
\end{figure}

  \subsection{Physical assumptions} \label{subsec:physical_assumptions}
  
For a full description of the code, we refer to the references cited at the start of this section. Here we discuss the main assumptions that are of direct relevance to this study.

\paragraph{Stellar lifetimes}-- The evolutionary tracks and stellar lifetimes (until reaching the white dwarf phase or core collapse) of single stars in our simulations originate from the grid of detailed non-rotating stellar evolutionary models of \citet{Pols+1998} computed with an updated version of the STARS code \citep{Eggleton1971a,Eggleton1972a,Pols+1995}.  For stars up to 20\Msun , we use the fitting formulae of \citet{Hurley+2000} for these models. At higher mass, we switch to a logarithmic tabular interpolation of the lifetimes by \citet{Pols+1998}, and above 50\Msun , we extrapolate as described in \citet{Schneider+2015}.  

Our resulting mass-lifetime relation is shown in \autoref{fig:lifetime}. We find good agreement with  simulations with the evolutionary code MESA, version 7184 \citep{Paxton+2011,Paxton+2013,Paxton+2015} for non-rotating stars when using our standard metallicity, $Z=0.014$, and the Schwarzschild criterion for convection with a step-overshooting parameter $\alpha_{ov} = 0.335 H_p$, where $H_p$ is the pressure scale height, as calibrated by \citet{Brott+2011}.  

The widely-used Geneva models of \citet{Schaller+1992} predict lifetimes that are 10-15\% shorter, as can be seen in \autoref{fig:lifetime}. These are models of non-rotating stars with a metallicity $Z=0.02$. Our MESA simulations give similar lifetimes when assuming no overshooting, $\alpha_{\rm{ov}} = 0$. Given the evidence for extra mixing processes beyond the convective core based on calibrations of the overshooting parameter $\alpha_{\rm{ov}}$ \citep[e.g.,][]{Ribas+2000,Claret2007,Brott+2011}, we consider the lifetime predictions by our models with overshooting to be more realistic.

\paragraph{Stellar winds}-- We include updated  mass-loss prescriptions as described in \citet{de-Mink+2013}, which include the recipes of \citet{Vink+2000}. At luminosities in excess of $4000\Lsun$, we switch to the empirical mass-loss rates of \citet{Nieuwenhuijzen+1990} when these rates exceed those by \citet{Vink+2000}.  To account for the empirical boundary of stars in the upper part of the Hertzsprung-Russell diagram as described in \citet{Humphreys+1994}, we add a factor in the mass loss as described in \citet{Hurley+2000} to simulate the enhanced mass loss of Luminous Blue Variables (LBV) that are thought to reside near this boundary. For stars that are stripped from their hydrogen envelopes, we adopt the Wolf-Rayet (WR) mass-loss prescription by \citet{Hamann+1995} and \citet{Hamann+1998} reduced by a factor of $10$ to account for the effect of wind clumping \citep{Yoon2015}.  For post-main-sequence stars, Asymptotic Giant Branch (AGB) stars, and thermally pulsating AGB stars, we use \citet{Kudritzki+1978}, \citet{Vassiliadis+1993}, and \citet{Karakas+2002} respectively, as described in \citet{Izzard+2009}.

Our mass-loss prescriptions scale with metallicity as $\dot{M} \propto (Z/Z_\odot)^m$ where $m = 0.69$ in main-sequence stars \citep{Vink+2001, Mokiem+2007}. In post-main-sequence phases, we adopt $m = 0.5$  \citep{Kudritzki+1989}.
In the WR phase, mass loss scales with metallicity assuming a power-law index of $0.86$ \citep{Vink+2005}. In the LBV phase, mass loss is assumed to be invariant for metallicity.  

The mass-loss rate by stellar winds as well as eruptive events is uncertain, in particular for the late phases and the most massive stars  \citep{Smith2014}. In the mass range we are most interested in, mass loss during the late phases only affects the stellar envelope. It does not have a large impact on the core of the stars and thus on the remaining lifetimes, which are the main focus of this work. Nevertheless, we explore the impact of changes in the mass-loss rate by multiplying the mass loss by an efficiency factor, $\eta$, which we set to unity in our standard simulations.  We consider the variations $\eta = 3$  and $\eta = 0.33$ for all mass-loss rates simultaneously.

\paragraph{Tides}-- We account for the effect of tides on the stellar spins and the stellar orbits of stars in binary systems \citep{Zahn1977, Hurley+2002} and the transfer of angular momentum during mass transfer via an accretion disk or the direct impact of the accretion stream onto the surface as described in \citet{de-Mink+2013} following \citet{Ulrich+1976} and \citet{Packet1981}. We assume that the stellar spins are aligned with the orbit \citep{Hut1981}.

\paragraph{Mass transfer}-- When a star fills its Roche lobe and mass transfer is stable, we compute the mass-loss rate from the donor star by removing as much mass as needed for the star to remain inside its Roche lobe. The resulting mass transfer rates are capped by the thermal timescale of the donor. We define the mass transfer efficiency, $\beta,$ as the fraction of the mass lost by the donor that is accreted by the companion, 
\begin{equation}
 \beta  \equiv \, \left| \frac{\dot{M}_{\rm acc}}{ \dot{M}_{\rm don}} \right|.
\end{equation}
If mass is transferred on a timescale that is much shorter than the thermal timescale of the accreting star $\tau_{\rm th,acc} $, the star will be driven out of thermal equilibrium and expand \citep{Neo+1977}.  Although our simulations do not follow this phase in detail, it is expected that the companion can only accrete a fraction of the transferred material when  $ \lvert \dot{M}_{\rm don} \rvert \gg   \dot{M}_{\rm acc, th} \equiv M_{\rm acc} /  \tau_{\rm th,acc}  $, where $M_{\rm acc}$ is the mass of the accreting star. In line with this physical picture, we limit the mass accretion rate to 

\begin{equation}\label{eq:beta}
\lvert \dot{M}_{\rm acc} \rvert  = \min \left(\lvert \dot{M}_{\rm don} \rvert, \,\, f \frac{ M_{\rm acc} }{ \tau_{\rm th,acc} }\right), 
\end{equation}
where $f$ is an efficiency parameter for which we adopt 10 in our standard simulation (to reproduce the mass transfer efficiency of  \citealt{Schneider+2015}).  The mass transfer efficiency in this case,  which we refer to as $\beta_{\rm th}$, varies between 0 and 1 depending on the physical properties of the donor and the accretor. The efficiency of mass transfer is poorly constrained  \citep[see, e.g., the discussion in][]{de-Mink+2007}.  We therefore also explore the extreme case where none of the transferred mass is accreted,  $\beta=0$, the case of very inefficient mass accretion, $\beta=0.2$, as well as fully conservative mass transfer, $\beta=1$.

\paragraph{Angular momentum loss}-- Mass that is lost from the system also takes away angular momentum. The specific angular momentum $h$, carried away from the system during mass loss, is parametrized by,

\begin{equation}
h = \gamma \frac {J_{\rm orb}} {  M_{\rm acc} + M_{\rm don} },
\end{equation}
where $J_{\rm orb}$ is the total orbital angular momentum, $M_{\rm acc}$ and $M_{\rm don}$ are the masses of the accretor and donor star respectively and $\gamma$ is a free parameter. 
In our standard simulation, we assume that mass lost from the system is emitted in a spherical wind or bipolar outflow originating from the accreting star \citep{van-den-Heuvel1994}. Thus, the specific angular momentum, $h,$ that the lost mass carries is equal to the specific orbital angular momentum of the accreting star, which yields $\gamma = \gamma_{\rm orb,acc} \equiv M_{\rm don} / M_{\rm acc}$.

We also consider the extreme limiting case of negligible angular momentum transported by the mass lost from the system during mass transfer ($\gamma = 0$). We further consider the case where mass is lost through the outer Lagrangian point, forming a circumbinary disk. Based on simulations by \citet{Artymowicz+1994}, we consider that the binary system will clear out the inner portion of the disk by resonance torques. We explore the case that an inner region of size $r_{\min} = 2a$ is cleared, where $r_{\min}$ is the inner radius of the circumbinary disk and $a$ the separation of the binary system.  This is consistent with typical values found by  \citet{Artymowicz+1994}.  We thus consider the case of  $\gamma = \gamma_{\rm disk}$,
\begin{equation}
\gamma_{\rm disk}  \equiv \frac{ (M_{\rm acc} + M_{\rm don})^2}{ M_{\rm acc} M_{\rm don}} \sqrt{\frac{r_{\min}}{a} }. 
\end{equation}

\paragraph{Contact and common envelope evolution}-- To decide which systems come into contact or experience common envelope (CE) evolution,  we consider a critical mass ratio, $q_{\rm crit}$.   
In binary systems with a post-main-sequence donor, we assume that systems with  $M_{\rm acc}/M_{\rm don} < q_{\rm crit}$ enter a common envelope phase.  We follow the prescriptions of \citet{Hurley+2002} for $q_{\rm crit}$, except for the case when the donor fills its Roche lobe while experiencing hydrogen shell burning and crossing the Hertzsprung gap (HG), where we use $q_{\rm crit, HG} = 0.4$  \citep{de-Mink+2013}. We use the same value for the naked helium star donors that experience helium shell burning (HeHG).

Common envelope evolution may either lead to the removal of the envelope or, if the ejection is not successful, a merger.  In our treatment of common envelope evolution we use the formalism described in \citet{Tout+1997} based on  \citet{Webbink1984}, \citet{Livio+1988} and \citet{de-Kool1990}.  In this formalism, two parameters are introduced, the efficiency parameter of ejection, $\alpha_{\rm CE}$, and $\lambda_{\rm CE}$ which parametrizes the binding energy of the envelope \citep[see eq. 73 and 69 in][respectively]{Hurley+2002}.  In our standard model, we assume that $\alpha_{\rm CE}$ is unity \citep[e.g.,][]{Webbink1984, Iben+1984,Hurley+2002}, but we also run models with a range of values ($0.1$, $0.2$, $0.5$, $2$, $5$, $10$) to probe the large uncertainties associated with this phase of evolution. By varying the efficiency parameter, we also implicitly consider the effect of uncertainties in the binding energy $\lambda_{\rm CE}$ as $\alpha_{\rm CE} \lambda_{\rm CE}$ appears as a product in the expression. Values of $\alpha_{\rm CE} > 1$ account for possible extra energy sources used to unbind the envelope apart from the orbital energy \citep[e.g.,][]{De-Marco+2011,Ivanova+2016}. To compute the envelope binding energy parameter, $\lambda_{\rm CE}$, we use fits to detailed models \citep{Dewi+2000,Dewi+2001,Tauris+2001}. We also consider a model variation where we adopt a constant value $\lambda_{\rm CE}=0.5$ \citep[e.g.,][]{de-Kool1990}.  For a discussion of the limitations of this formalism we refer to \citet{Ivanova+2013}. 

In systems with a main-sequence donor, we adopt $q_{\rm crit, MS} = 0.65$ to account for systems that come into contact during the rapid thermal timescale mass transfer phase, which is consistent with the detailed models by \citet{de-Mink+2007}. Alternatively, binary systems may come into contact because of their own nuclear timescale evolution. In main-sequence stars, we assume that contact leads to a merger. 

\paragraph{Mergers and rejuvenation}--
In case a merger occurs, we follow Table 2 of \citet{Hurley+2002} to determine the outcome. When two main-sequence stars (MS+MS) merge, we follow the updated algorithm by  \citet{de-Mink+2013,de-Mink+2014}, based on \citet{Glebbeek+2013}, to account for mass loss and internal mixing. This algorithm uses two parameters,  $\mu_{\rm loss} = 0.1,$ which is the fraction of the total mass lost from the system during the merger, and $\mu_{\rm mix} = 0.1,$ which is the fraction of the remaining envelope mass that is mixed into the convective core. The values above are adopted in our standard simulation. We vary $\mu_{\rm loss}$ between 0 and 0.25 and $\mu_{\rm mix}$ between 0 and 1.  

We account for the rejuvenating effect of mixing of fresh hydrogen into the central regions of accreting stars and mergers.  For this, we use fits to the effective mass of the convective core by \citet{Glebbeek+2008a} as described in \citet{de-Mink+2013} and \citet{Schneider+2015}.

\paragraph{Minimum mass for core-collapse supernovae}--
We predict the final fate of the stars in our simulation using our estimate of the final metal core mass, that is, the mass of the core consisting of elements heavier than helium, sometimes referred to as carbon oxygen (CO) core mass.  We adopt a minimum metal core threshold of $M_{\min, \rm metal} = 1.37 \Msun$ for a collapse \citep{Nomoto1984,Nomoto1987,Podsiadlowski+2004,Takahashi+2013}. In our standard simulations, this corresponds to a minimum single star initial mass of $M_{\min, \rm cc} = 7.53\Msun$.  We also consider the variations $ M_{\min, \rm metal} = 1.30$ and $1.40\Msun$ when exploring the sensitivity of our results. In single stars, these values correspond to a minimum initial mass of  $M_{\min, \rm cc}\approx7$ and $8$ $\Msun$ , respectively, in our standard metallicity.  
We do not explicitly distinguish between electron capture and iron ccSN \citep[e.g.,][]{Tauris+2015}. We also do not consider the accretion-induced-collapse of white dwarfs.

It is uncertain whether or not the collapse of the core leads to a successful supernova explosion in all cases and whether or not this explosion is a bright event. Especially in massive stars, the explosion may fail to eject the outer layers resulting in fall back of material. This possibly leads to fainter explosions \citep[e.g.,][]{OConnor+2011,Ugliano+2012,Nadezhin1980, Lovegrove+2013, Piro2013}, or even the simple disappearance of a star without any electromagnetic signature \citep[e.g.,][]{Kochanek+2008}. In our standard simulations, we assume that all core collapses result in an observable event. We also run model variations in which we exclude all supernovae from stars with metal cores more massive than a single star of $M_{\max, \rm cc} = 35\Msun$ and $M_{\max, \rm cc} = 20\Msun$ would produce, mimicking, in a simplified way, the effect of `failed' explosions.

\paragraph {Supernova kick}--
We account for the effect of sudden mass loss during the supernova explosion on the orbit  \citep{Blaauw1961, Boersma1961, Hurley+2002}. In addition, at the onset of the supernova, asymmetries in the explosion mechanism may lead to a natal kick to the compact remnant. We assume a natal kick for the compact object as in equation (A15) of \citet{Hurley+2002}, where the remnant receives a kick in a random direction and with a scalar velocity drawn from a 1-D Maxwellian distribution characterized by a 1-D root mean square of $\sigma = 265\, \mathrm{ km/s}$ \citep{Hobbs+2005}. We also examine the effect of the extreme cases with no supernova natal kick at the remnant ($\sigma_0$) and with kicks strong enough to always disrupt a binary system after the explosion.

 \begin{table*}[ht]
  \caption{Summary of the key parameters adopted in our standard simulations and the variations that we consider. See \autoref{Section:Model} for a description of the symbols and further assumptions. 
  \label{table:parameters_standard}}
  \centering
   \small
\begin{tabular}{lllcc}
  \hline \hline 
     \\[0.1ex]
&& Symbol  & Standard models$^{(a)}$  & Model variations
 \\[0.1ex]
\hline
   \\
\multicolumn{5}{l}{\bf Physical assumptions} \\
- & mass transfer efficiency &  $\beta$ &                              $\beta_{\rm th}^{}$&  0, 0.2, 1\\  
- & angular momentum loss  & $\gamma$  &             $\gamma_{\rm orb,acc}$ & 0, $\gamma_{\rm disk}$ \\
- & mass loss during merger of two MS stars &  $\mu_{\mathrm{loss}}$                  &  $0.1$ & 0, 0.25 \\ 
- & mixing during merger of two MS stars & $\mu_{\mathrm{mix}}$                         &  $0.1$ & 0, 1\\  
- & natal kick compact remnant  (km\,s$^{-1}$)  &  $\sigma$                        &265 & $\sigma_0$, $\infty$  \\
- & common envelope efficiency &  $\alpha_{\mathrm{CE}} $                       &  $ 1$  &  0.1, 0.2, 0.5, 2, 5, 10\\
- & envelope binding energy &  $\lambda_{\mathrm{CE}} $                           & $\lambda_{\rm Dewi+00}$  & 0.5 \\
- & critical mass ratio for contact for MS donor  &     $q_{\mathrm{crit,MS}}$              &  $ 0.65$ & 0.25, 0.8 \\
- & critical mass ratio for unstable mass transfer for HG donor & $q_{\mathrm{crit,HG}}$ &   $ 0.4 $ & 0, 0.25, 0.8, 1\\
- & stellar-wind mass-loss efficiency parameter & $\eta$ &  1 & 0.33, 3 \\
- & maximum single-star equivalent birth mass for ccSN (\Msun) & $M_{\max, \rm cc}$ &  100 & 20, 35\\
- & minimum metal core for ccSN (\Msun) & $M_{\min, \rm metal}$ &  1.37 & 1.3, 1.4\\

  \\
\multicolumn{5}{l}{\bf Initial conditions} \\
- & slope initial mass function & $\alpha$              &  $ -2.3 $ &  -1.6, -2.7, -3.0 \\
- & slope initial mass ratio distribution & $\kappa$   &   $ 0 $  & -1, 1 \\
- & slope of initial period distr.   & $\pi$  &      $\pi_{\rm Opik24,Sana+12}$  & -1, 1 \\
- & metallicity & $Z$ &   $0.014$  &  0.0002, 0.004, 0.008, 0.02, 0.03 \\
- & binary fraction$^{(a)}$  & $f_{\mathrm{bin}}$                              &  $0.7$ , $0.0^{(a)}$ & 0.3, 1, $f_{\mathrm{bin}} (M_1)^{}$ \\  
- & normalization parameter  (\Msun)  & $M_{\mathrm{low}}$      &  $ 2 $ & 1, 3 \\
  \hline
  \end{tabular}
\tablefoot{
\tablefoottext{a}{The difference between our two standard models is that in one we simulate only single stars and in the other we assume a binary fraction of $0.7$.}\\
}  
\end{table*}

\subsection{Simulation set up}\label{subsec:simulation_setup}

In our standard assumptions, we evolve $10^{4}$ single stars and more than $3\times10^6$ binary systems with varying primary masses, mass ratios, and orbital periods on a grid of $150\times150\times150$ systems. Test simulations indicate that these resolutions are sufficient for the purpose of this work. The difference between our two standard models is that in one we take into account only single stars and in the other we assume a binary fraction of $0.7$. We take primary masses spaced at equal logarithmic intervals between $M_1=3$ and $100$\Msun. The lower limit encompasses all systems in our simulations  with the potential to result in a core-collapse event within a safe margin.  We take mass ratios linearly spaced between $q=0.1$ and $1$ and orbital periods spaced at equal logarithmic intervals between  $\log_{10} P (d)  = 0.15$ and $3.5$.  We weigh each system according to the initial distribution functions specified in \autoref{subsec:initial_conditions}.  When computing variations in the assumptions, we reduce the resolution of the grids in each dimension by a factor of two resulting in approximately  $4.2 \times 10^5$ systems in each model variation.

\section{Results}\label{Section:Results}

The lifetime of a star is determined by the amount of nuclear fuel available and the rate at which it burns this fuel. In single stars, both of these properties are primarily a function of the initial mass, as shown in  \autoref{fig:lifetime}.  
The lifetime approximately scales with mass as $\tau(M)\approx M^{-x}$.  The relation is steeper for intermediate-mass stars ($x \approx 2.4$ near $M = 5\Msun$) and flattens at higher masses ($x \approx 0.6$ near $M= 50\Msun$).  At the highest masses, the lifetime converges to a finite value of approximately 3\Myr. For stars in binary systems, the lifetime is no longer a simple function of their initial mass alone, due to possible mass transfer or merging.

In this section, we discuss the distribution of stellar lifetimes of single stars and binary systems that produce a core-collapse supernova. The lifetime sets the time-delay between formation of the star and the moment its core collapses. We therefore refer to this distribution as the \emph{delay-time distribution} (DTD) of ccSNe.  We first discuss the case of a population of single stars in \autoref{sec:DTD_single}, followed by the case of a population that contains a realistic fraction of binary systems in \autoref{sec:DTD_realistic}.

\begin{figure}[t]\center
\includegraphics[width=0.5\textwidth]{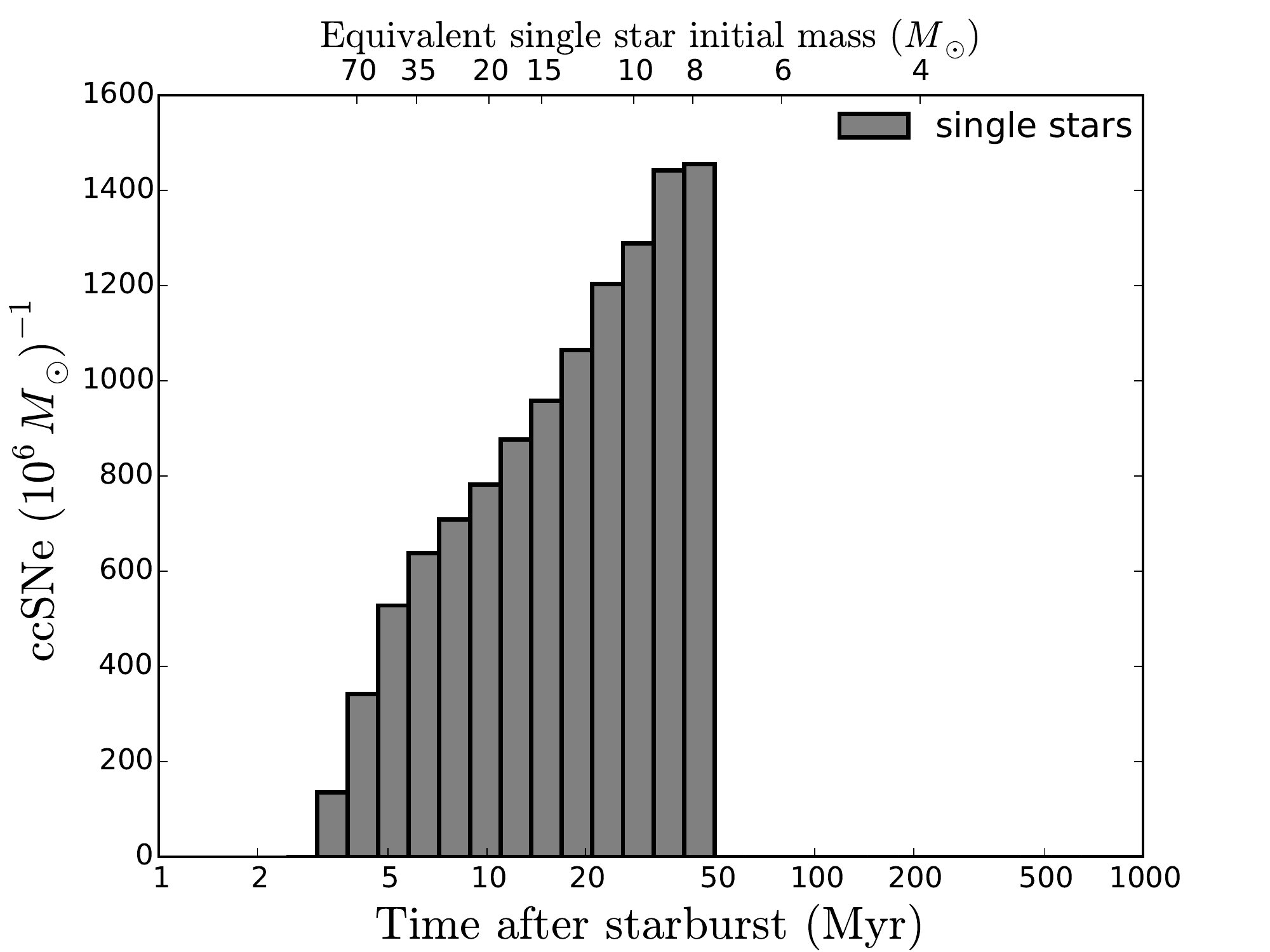}
\caption{The delay-time distribution of core-collapse supernovae based on our single star models.  The diagram shows  the number of events per logarithmic time bin for a starburst of $10^6\Msun$ in our standard model. The top axis shows the initial mass of single stars with the corresponding lifetime given in the bottom axis, computed with binary\_c.  The most massive stars evolve most rapidly and end their life after approximately $ 3$\Myr. There are no core-collapse events after approximately $48$\Myr when the least massive single star that can undergo core collapse explodes. \label{fig:SN_delay_time_single}}
\end{figure}

\subsection{Delay-time distribution of single stars \label{sec:DTD_single}}

In a population of single stars formed in an instantaneous starburst at $\tau=0,$ we expect ccSNe between  $\tau_{\min, \rm cc} \approx 3\Myrs$, corresponding to the lifetime of the most massive star in our simulation, and  $\tau_{\max, \rm cc} \approx 48\Myr$, corresponding to the lifetime of the least massive star that undergoes core collapse in our simulation, with $M_{\min, \rm cc} = 7.53\Msun$ (\autoref{fig:lifetime}). 
This can be seen in  \autoref{fig:SN_delay_time_single} where we show the DTD for single stars for a $10^6\Msun$ starburst.  The diagram shows the number of ccSNe per logarithmic time bin. 

The distribution rises steadily between  $\tau_{\min, \rm cc}$ and  $\tau_{\max, \rm cc}$. The slope of the distribution is dictated by the slope of the IMF and the derivative of $\tau^{-1}(M)$, the inverse of the lifetime-mass relation. The IMF favors less massive stars that contribute in the later time bins. The flattening of the lifetime-mass relation for the highest masses leads to a relative pile-up of ccSNe in the early time bins. The net effect is a distribution that rises steadily, when expressed in the units chosen here, with a slope that is slightly steeper at early times, $\tau \lesssim 5\Myr$.  The average delay time of the distribution is $\tau_{\rm{av}} = 17.5 \Myrs $, the median is $\tau_{\rm{50\%,\rm{all}}} = 19.6 \Myrs $.

The units and axes used here have the advantage that an equal number of systems occupy an equal area in this diagram, which is useful as a visualization of the discussion in the paragraphs that follow. For convenience, we also show the evolution of the ccSN rate (the number of events \emph{per year} per mass) in \autoref{Appendix:fit}. The rate for single stars peaks around 5.1\Myrs in our simulation to a value of approximately $4.75 \times 10^{-10}$ events per yr per $\Msun$, as shown in \autoref{fig:SN_delay_time_snrate}.

\subsection{Delay-time distribution including binaries \label{sec:DTD_realistic}}

\begin{figure*}[t] \center
\includegraphics[width=0.85\textwidth]{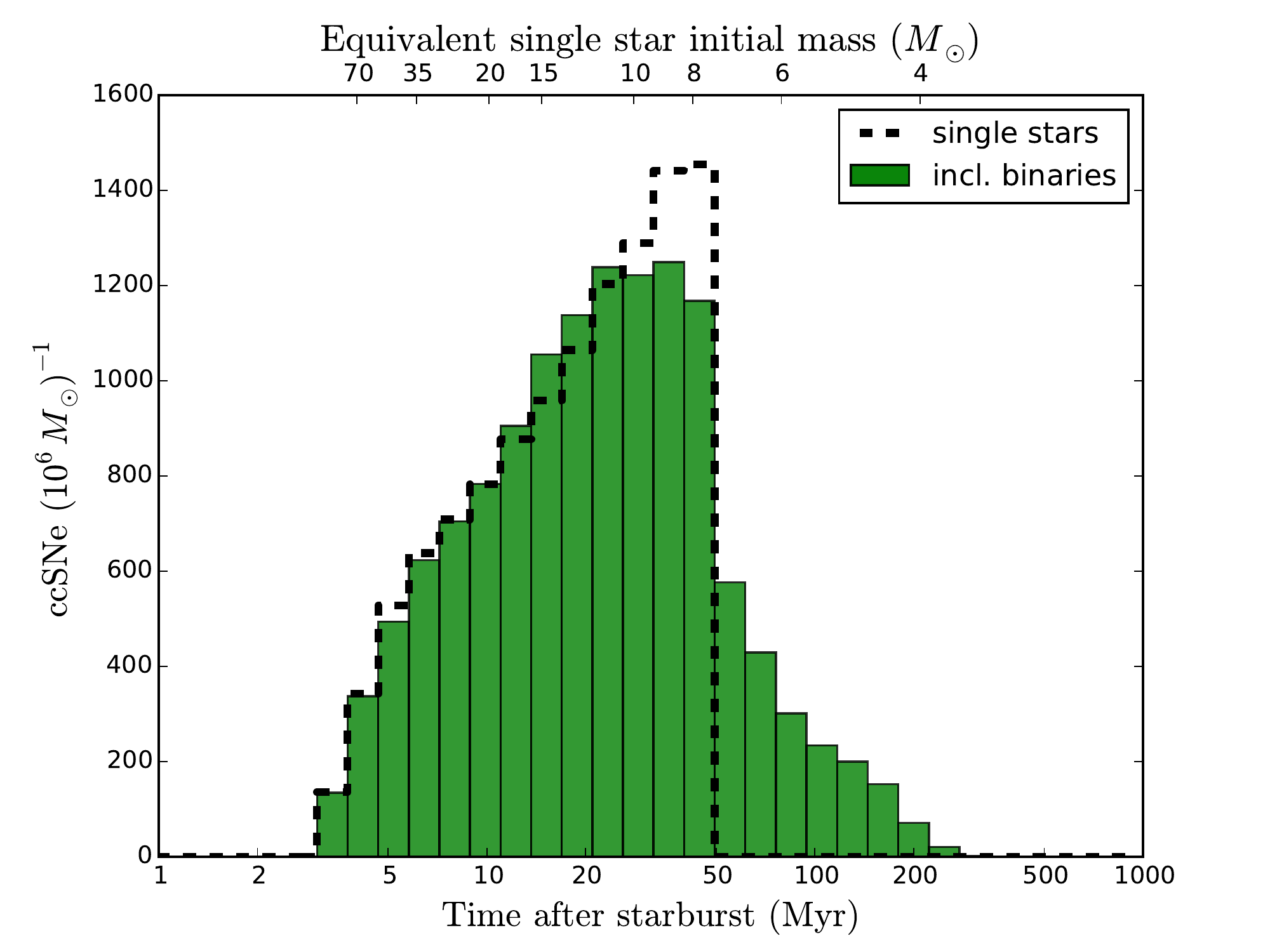}
\caption{The delay-time distribution of core-collapse supernovae for a population consisting of $70\%$ binary systems (green histogram) compared to the distribution for a population of only single stars (black dashed line). It shows  the number of events per logarithmic time bin for a starburst of $10^6\Msun$ for our standard models. The top axis shows the initial mass of single stars with the corresponding lifetime given in the bottom axis, computed with binary\_c. The most striking difference is the fraction of `late' core-collapse supernovae ($f_{\rm late} = 15.5^{+8.8}_{-8.3} \%$), after the last massive single star explodes at $\tau_{\max, \rm cc} \approx 48$ \Myrs. The errors in the fraction above result from variations of our standard assumptions.  \label{fig:SN_delay_time}} 
\end{figure*}

The picture changes when we account for a realistic fraction of binary systems.  Some  intermediate-mass stars, with $M< M_{\min, \rm cc}$, can accrete mass from their companions and become massive enough to experience the advanced nuclear burning stages. As a result, they end their lives as supernovae leaving neutron stars as remnants instead of becoming white dwarfs. Mass accretion also rejuvenates a star because fresh fuel is mixed into the central burning regions, effectively prolonging its life.  However, mass accretion also accelerates aging. Making the star more massive increases its luminosity and thus the rate at which it burns its remaining fuel.  Conversely, mass stripping can, under certain conditions, prevent a massive star from ending its life as a supernova. This is only true if mass loss occurs early in its evolution, before the star has a fully developed core.  We investigate how all these binary evolution processes interplay and compete in affecting the DTD.  

In \autoref{fig:SN_delay_time} we show the distribution of ccSNe resulting from a $10^6\Msun$ starburst event in our standard simulation that accounts for binaries. The single star DTD is over-plotted for comparison.  Both distributions are remarkably similar at early times. The differences become evident at around 20\Myrs, where the binary distribution peaks.  At around 30-50\Myrs, the binary distribution shows a deficit with respect to the single star distribution. The deficit results from close systems in which the primary star has a mass just above but close to $M_{\min, \rm cc}$. If it is stripped of its envelope early in its evolution, that is, before the completion of hydrogen burning, this can prevent its core from growing enough in mass to reach the advanced burning stages. Thus, interaction with a companion prevents these stars from ending their lives as ccSNe. 

The most striking feature that distinguishes the binary star distribution from that of single stars is the prominent excess of events between 50 and 200\Myrs. These occur after the last single star exploded at $\tau_{\max, \rm cc} \approx 48\Myr$. We refer to them as ``late core-collapse supernovae'' and discuss them in detail in \autoref{sec:properties_of_late_ccsne}.  These late events account for  $15.5^{+8.8}_{-8.3} \%$ of the total number of ccSNe.  The main value quoted here corresponds to the relative contribution of late events with respect to all core-collapse events as found in our standard simulation. The errors reflect the minimum and maximum we find when considering model variations, as we discuss in section \ref{sec:robustness}. The above ratio of late ccSNe to total ccSNe will be referred to as $f_{\rm late}$ from now on.

The average delay time for a population including a realistic fraction of binaries is  $\tau_{\rm{av}} = 21.6$\Myrs, around 20\% longer than in our single star simulation.  The median time delay is  $\tau_{\rm{50\%,\rm{all}}} = 22.1\Myr$, nearly 10\% longer than in our single-star simulation. 

We observe a tail of very late events with delay times $>300$\Myr. They are extremely rare in our standard simulation accounting for less than 0.01\% of all core-collapse events. However, in some of the variations that we consider, we find higher fractions of 0.03\%.  These very late events result from a variety of rare binary evolutionary paths. They typically originate from systems with a low-mass secondary ($<4 \Msun$) in an initially wide orbit ($\sim 1000 $ days). The systems usually evolve through multiple CE and mass transfer episodes, eventually leading to a merger of a massive white dwarf (the primary) with a helium star secondary. Some of these may result in an accretion-induced collapse, but to determine the fate of these mergers requires detailed modeling beyond the scope of this work.

\subsection{Total number of core-collapse supernovae \label{totalNumber}}

Accounting for binaries increases the number of ccSNe for the same total stellar mass of a population. We find a relative increase of  $14^{+15}_{-14}$\% when accounting for a realistic binary population. The increase is due to the added contribution of intermediate-mass systems, which are favored by the IMF.  However, the increase is limited by the fact that we normalize our simulations by the total stellar mass; if the average mass of single stars is $\langle M \rangle$ then that of a binary systems is $\langle M_{\rm 1} + M_{\rm 2} \rangle = \langle M  \rangle \, + \, \nicefrac{1}{2}\, \langle M \rangle  = \nicefrac{3}{2}\, \langle M \rangle$, assuming that single and primary stars follow the same IMF and taking into account that the typical mass ratio is $q \approx0.5$. Thus, in a population of the same total mass, the number of systems is lower when binary systems are included. We find that the net result of both the above effects is an increase in the number of ccSNe in a population that contains binary systems. 

The total number ${\cal N}$ of ccSNe in a $10^6\Msun$ stellar population is less well constrained because of uncertainties in the normalization of the mass contained in low-mass stars. In our standard simulations, we find  $ {\cal N } = 1.14 \times10^4$ for single stars and ${\cal N } = 1.30 \times10^4$ for a realistic binary fraction. Variations in the IMF slope alone ($\alpha=-3.0$ and $-1.6$) lead to  $ {\cal N} =  0.28-2.53 \times10^4$, respectively, in the single-star population  and $ {\cal N} = 0.35-2.53 \times10^4$  in the simulation that includes binaries.  All other model variations considered in \autoref{table:parameters_uncertainties} lead to changes in the total number of events by less than 25\%.

\begin{figure*}[t]\center
\includegraphics[width=\textwidth]{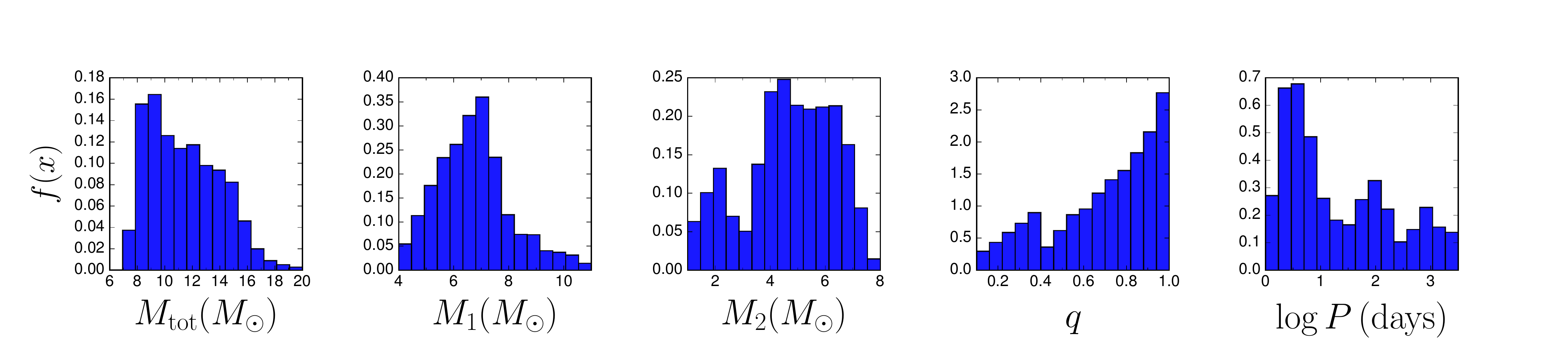}\caption{ The progenitor properties of `late' core-collapse supernovae for our standard simulation with binary stars. The normalized histograms show the distribution of the initial total mass $M_{\rm tot}$, the initial mass of the primary, $M_1$, and secondary, $M_2$, the initial mass ratio, $q \equiv M_2/M_1$, and the initial orbital period $P$ of binary systems that produce at least one core collapse after the last single star has exploded at around $48$ \Myrs .  
\label{fig:progenitors}}
\end{figure*}

\section{Late core-collapse supernovae and their progenitors}\label{sec:properties_of_late_ccsne}

Late ccSNe occurring around 50-200\Myr after star formation are not predicted in single stellar evolution. They are exclusively the product of interacting binaries and, as mentioned, they account for $f_{\rm late} = 15.5^{+8.8}_{-8.3} \%$ of all ccSNe.
Almost all systems that lead to a late event have at least one intermediate-mass star with $M <  M_{\min, \rm cc}  = 7.53\Msun$, the threshold mass for a single star to undergo core collapse. Approximately three out of four late events result from systems in which both stars have initial masses below \Mmincc. Both stars in these systems were destined to end their lives as white dwarfs, if it were not for the interaction with their companion.  This can be seen in \autoref{fig:progenitors}, where we show the normalized distributions of the initial properties of the progenitor systems. We provide the initial total mass, $M_{\rm tot}$, the initial mass of the primary and secondary, $M_1$ and $M_2$ respectively, the initial mass ratio, $q \equiv M_2/M_1$, and the initial orbital period $\log_{\rm 10} P$.  

The distribution of initial total mass for the progenitor systems peaks at 9\Msun (leftmost panel of \autoref{fig:progenitors}). It extends up to around 15\Msun, that is, approximately 2\Mmincc, rapidly declining for higher masses. The distribution drops off steeply for masses lower than the peak with a minimum approaching \Mmincc. 

The typical mass of the primary star is in the range 5--8\Msun (second panel of \autoref{fig:progenitors}).  Stars in this mass range take around 50-200\Myrs to evolve off the main sequence and start interacting. Hence, it is mainly the relatively slow evolution of the primary that causes the eventual explosions in these systems to be late. A secondary reason for longer delay times is rejuvenation of the accreting star (or merger product).

The distribution of secondary masses ranges from approximately 1 - 8 \Msun and peaks near 4.5\Msun, as can be seen in the third panel of \autoref{fig:progenitors}. There is a preference for systems that initially have mass ratios near unity, as shown in the fourth panel. Most progenitor systems originate from systems with initial orbital periods less than approximately $30$ days, as can be seen in the rightmost panel. Systems with initial mass ratios near unity and orbital periods below $\sim 30$ days generally experience a more conservative first phase of mass transfer in our simulations \citep[cf. Figure~3 of][]{Schneider+2015}. This means that a significant fraction of the mass lost by the primary star is accreted by the secondary and is thus retained in the system. This is necessary for an intermediate-mass binary system to produce a ccSN because in most cases neither of the stars is individually massive enough to produce a collapse.    

\begin{figure}[t]\center
\includegraphics[width=0.5\textwidth]{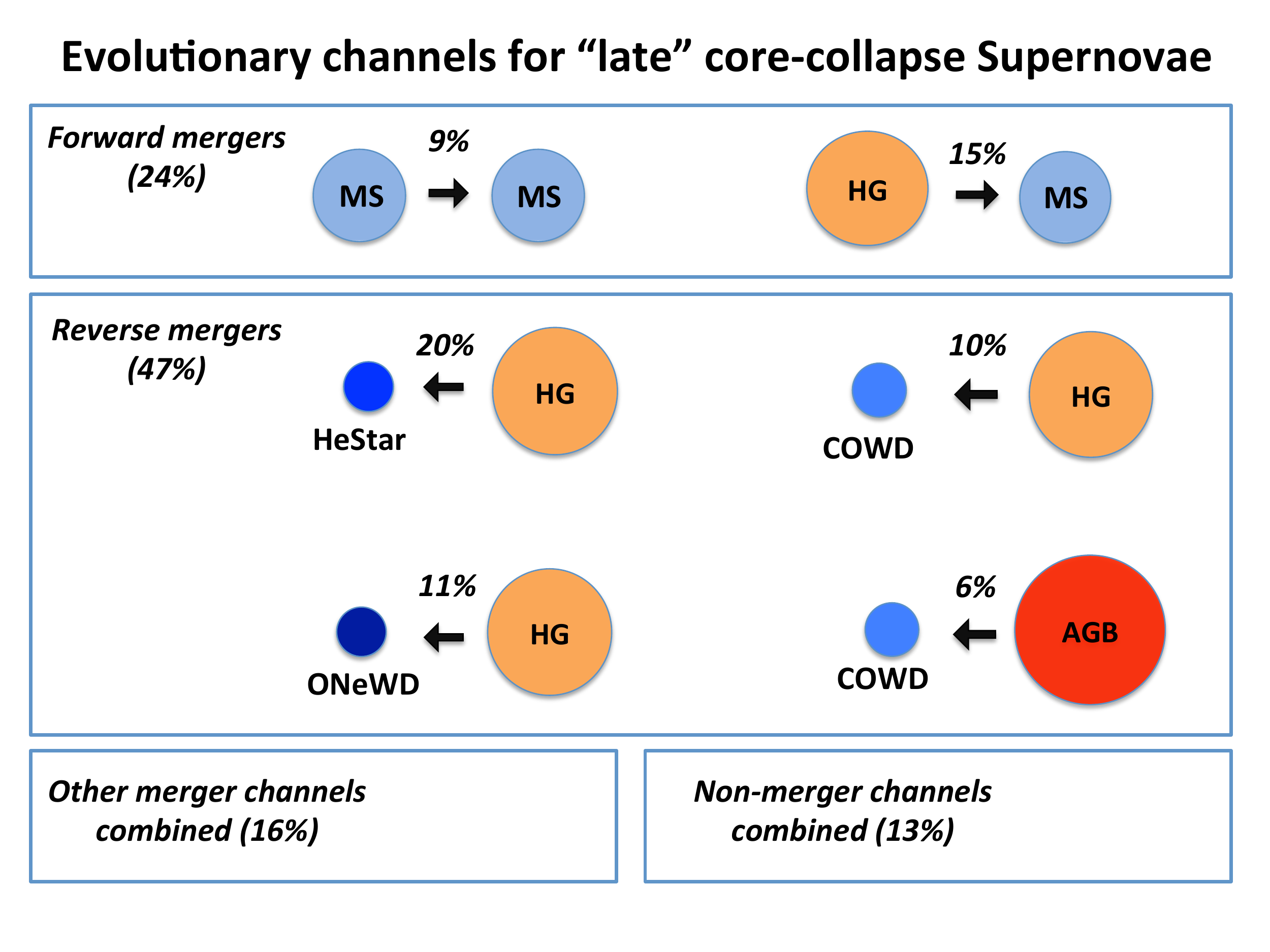}
\caption{ 
The main channels for `late' core-collapse supernovae in our standard simulation and their relative contribution to the number of late events (only showing those that contribute $>$5\% of the number of late ccSNe, i.e., $>$0.75\% of the total number of ccSNe).  For merger channels, we indicate the evolutionary state of the primary and secondary star at the onset of merging. Arrows indicate the direction of mass transfer preceding the merger, which can be forward (from primary to secondary) or reverse (from secondary to primary). MS: main sequence, HG: Hertzsprung gap, COWD and ONeWD: carbon-oxygen and oxygen-neon-magnesium white dwarf,  HeStar: naked helium star, AGB: Asymptotic Giant Branch. \label{table:channels} }
\end{figure}

\begin{table*}[p!]
  \caption{Variations of the physical assumptions and initial conditions considered and their impact on  the fraction of late events, $f_{\mathrm late}$, with respect to the total number of core-collapse supernovae. We also provide the median delay time of late events, $t_{\mathrm {50\%,late}}$, and the delay time before which 90\% the late events are found, $t_{\mathrm {90\%,late}}$. Standard assumptions are listed in \autoref{table:parameters_standard} and symbols are explained in \autoref{Section:Model} \label{table:parameters_uncertainties}.}
  \centering
   {\small
  \begin{tabular}{llcrrr}
  \hline \hline 
   \\[0.01ex]
Model &Description & & $f_{\mathrm {late}}$ & $t_{\mathrm {50\%,late}} $& $t_{\mathrm {90\%,late}} $ \\
    &&&&(Myr) & (Myr)  \\
\hline
 \\[0.5ex]
{\bf 00} &  {\bf Standard Simulations } &                                &  {\bf  15.5\% } &   {\bf    71.9 } &  {\bf   150.5 }  \\
  \\
\multicolumn{5}{l}{\it \bf Physical assumptions} \\
 \\[0.1ex]
01 &  mass transfer efficiency& $\beta=0$                        &  $     8.8\%$ &     72.7 &    143.5  \\
02 &  $\quad$"" & $\beta=0.2$                      &  $    10.5\%$ &     61.1 &    140.9  \\
03 &   $\quad$""  &$\beta=1$                        &  $    21.7\%$ &     79.1 &    146.1  \\
04 &  angular momentum loss  &$\gamma=0$                      &  $    15.3\%$ &     70.6 &    146.1  \\
05 &  $\quad$""  & $\gamma= \gamma_{\rm disk}$      &  $    17.1\%$ &     71.4 &    130.3  \\
06 &  mass loss during merger of two MS stars & $\mu_{\mathrm{loss}}=0$                        &  $    14.9\%$ &     78.2 &    162.8  \\
07 &  $\quad$""  & $\mu_{\mathrm{loss}}=0.25$                     &  $    14.9\%$ &     70.6 &    142.6  \\
08 &  mixing during merger of two MS stars & $\mu_{\mathrm{mix}}=0$                         &  $    15.4\%$ &     72.3 &    148.7  \\
09 &  $\quad$""  & $\mu_{\mathrm{mix}}=1$                         &  $    15.8\%$ &     73.2 &    148.7  \\
10 &  natal kick compact remnant & $\sigma = \sigma_0$            &  $    16.3\%$ &     75.4 &    152.3  \\
11 &  $\quad$""  & $\sigma=\infty$       &  $    15.6\%$ &     72.3 &    147.8  \\
12 & common envelope efficiency  & $\alpha_{\mathrm{CE}} = 0.1$                          &  $    19.4\%$ &     77.7 &    147.8  \\
13 & $\quad$""  & $\alpha_{\mathrm{CE}}= 0.2$                           &  $    20.4\%$ &     81.5 &    158.9  \\
14 &$\quad$""   & $\alpha_{\mathrm{CE}} = 0.5$                          &  $    16.5\%$ &     81.0 &    162.8  \\
15 & $\quad$""   &$\alpha_{\mathrm{CE}}= 2.0$                           &  $    13.1\%$ &     69.3 &    145.2  \\
16 & $\quad$""   &$\alpha_{\mathrm{CE}} = 5.0$                          &  $     8.3\%$ &     63.0 &    137.5  \\
17 & $\quad$""   &$\alpha_{\mathrm{CE}} = 10.0$                         &  $     7.2\%$ &     61.8 &    137.5  \\
18 &  envelope binding energy &$\lambda_{\mathrm{CE}}=0.5$                           &  $    17.0\%$ &     82.0 &    178.1  \\
19 & critical mass ratios for contact/unstable mass transfer &$q_{\mathrm{crit,MS}} = 0.25$                         &  $    15.5\%$ &     72.7 &    149.6  \\
20 & $\quad$""   &$q_{\mathrm{crit,MS}} = 0.8$                          &  $    15.3\%$ &     71.9 &    146.9  \\
21 & $\quad$""   &$q_{\mathrm{crit,HG}} = 0.0$                     &  $    15.6\%$ &     79.6 &    170.8  \\
22 & $\quad$""   &$q_{\mathrm{crit,HG}} = 0.25$                    &  $    14.4\%$ &     75.4 &    154.2  \\
23 & $\quad$""   &$q_{\mathrm{crit,HG}} = 0.8$                     &  $    19.7\%$ &     73.6 &    124.9  \\
24 & $\quad$""   &$q_{\mathrm{crit,HG}} = 0.99$                    &  $    22.3\%$ &     74.9 &    138.4  \\
25 &  stellar wind mass-loss efficiency &$ \eta= 0.33$                                  &  $    15.7\%$ &     74.0 &    157.0  \\
26 &  $\quad$""   &$\eta= 3.0$                                   &  $    15.7\%$ &     71.3 &    141.7  \\
27 &  exclusion of WD mergers if companion $< M_{\min, \rm cc}$ &         &  $    13.2\%$ &     68.9 &    138.4  \\
28 &  maximum single star equivalent  birth mass for ccSN & $M_{\max, \rm cc} =35$                        &  $    16.9\%$ &     72.7 &    148.7  \\
29 & $\quad$""   & $M_{\max, \rm cc} =20$                         &  $    20.4\%$ &     72.7 &    148.7  \\
30 & minimum metal core for ccSN & $M_{\min, \rm metal} = 1.30$   &  $    14.8\%$ &     83.5 &    165.7  \\
31 & $\quad$""  &$M_{\min, \rm metal} = 1.40$  &  $    16.2\%$ &     62.6 &    132.7  \\

  \\
\multicolumn{5}{l}{\it \bf Initial conditions} \\
 \\[0.1ex]
32 & initial mass function  &$\alpha =-1.6$                   &  $     8.2\%$ &     68.5 &    142.6  \\
33 &  $\quad$"" &$\alpha =-2.7$                   &  $    20.4\%$ &     74.9 &    153.3  \\
34 &  $\quad$""  &$\alpha =-3.0$                   &  $    24.3\%$ &     76.3 &    157.9  \\
35 &  initial mass ratio distribution& $\kappa=-1$                    &  $    12.1\%$ &     64.9 &    140.9  \\
36 &  $\quad$""  &$\kappa=+1$                    &  $    18.1\%$ &     76.3 &    151.4  \\
37 &  initial period distribution &  $\pi=+1$                  &  $    11.3\%$ &     69.3 &    146.9  \\
38 &  $\quad$""  &  $\pi=-1$                  &  $    23.2\%$ &     78.2 &    151.4  \\
39 & metallicity & $Z=0.0002$                                            &  $    12.6\%$ &     98.2 &    199.7  \\
40 & $\quad$""  & $Z=0.001$                                             &  $    11.8\%$ &     92.5 &    179.2  \\
41 &$\quad$""  & $Z=0.004$                                             &  $    12.9\%$ &     83.0 &    159.8  \\
42 &$\quad$""  & $Z=0.008$                                             &  $    14.1\%$ &     78.2 &    152.3  \\
43 & $\quad$""  &$Z=0.02$                                              &  $    14.8\%$ &     68.5 &    141.7  \\
44 & $\quad$""  &$Z=0.03$                                              &  $    14.7\%$ &     64.9 &    134.3  \\
45 &  binary fraction &$f_{\mathrm{bin}} = 0.3$                              &  $     7.7\%$ &     72.7 &    148.7  \\
46 & $\quad$""  &$f_{\mathrm{bin}} = 1.0$                              &  $    20.0\%$ &     72.7 &    148.7  \\
47 &  mass dependent binary fraction &   $f_{\mathrm{bin}} (M_1)$                     &  $    12.8\%$ &     68.9 &    138.4  \\
48 & normalization parameter &$M_{\mathrm{low}}= 1\Msun $       &  $    15.5\%$ &     72.7 &    148.7  \\
49 & $\quad$""  &$M_{\mathrm{low}}=3\Msun$        &  $    15.5\%$ &     72.7 &    148.7  \\

 \hline
  \end{tabular}
  }
\end{table*}

\subsection{ Evolutionary channels }\label{properties_of_late_ccsne}

The evolutionary channels that lead to late events often involve multiple phases of mass transfer. The majority of late events,  87\% in our standard simulation, result from binary channels in which the two stars coalesce. 
We distinguish between \emph{`forward'} and \emph{`reverse'} mergers depending on whether it is the primary, that is, the initially more massive star, or the secondary that fills its Roche lobe and initiates the merger process, respectively. Below, we elaborate on these merger channels and after we discuss the non-merger scenarios. We limit the discussion to channels that contribute at least 5\% of the late events (i.e., more than approximately 0.75\% of the total core-collapse rate).  The less frequent merger channels that we do not discuss are responsible for a combined contribution of $16\%$ of the late events. 

\autoref{table:channels} provides a summary of the relative rates contributed by the main channels.  While our simulations cannot reliably follow the details of the merger process and the interior structure of the newly formed stellar object, we can predict relative rates of the different channels and asses the evolutionary status at the onset of the merger. This information can guide future numerical experiments. 

\paragraph{Forward mergers -- \label{sec:early_mergers}}

In our standard simulations, we find that 24\% of the late ccSNe come from systems that experience a merger in which the expansion of the primary star beyond its Roche lobe initiates the merger process. We distinguish the following cases.

\begin{description}
\item[\bf MS$\Rightarrow$MS:]
The merger between two main-sequence stars contributes 9\% of the late ccSNe in our standard simulations; they result from intermediate-mass primaries with initial masses of 5--8\Msun in tight binaries with initial orbital periods that are typically shorter than approximately $3$ days.  
The main contribution comes from systems with initial mass ratios $q< 0.65$. In such systems, the initial rapid phase of mass transfer drives the accreting star out of thermal equilibrium causing it to swell up. Such systems come into contact during the initial rapid mass transfer phase, denoted as Case AR, with A referring to the fact that the donor star still resides on the main sequence and R referring to the rapid mass transfer phase in the classification by \citet{Kippenhahn+1967} and \citet{Nelson+2001}. 
A second contribution comes from systems with initial mass ratios $q \approx 1$ in periods of less than approximately $2$ days, in which both stars evolve on a similar timescale. The stars come into contact when the secondary star also fills its Roche lobe due to its own evolutionary expansion. This occurs during the slow nuclear  time scale mass transfer phase (denoted as Case AS).

The primary has typically already completed over 80\% of its main sequence evolution at the moment of coalescence.  The secondary is in a similar evolutionary phase in Case AS, whereas it  is quite unevolved in Case AR, having only completed 10--30\% of its main-sequence evolution.
In Case AR mergers, the more evolved primary star is the more massive component at the onset of the merger,  at $5-8\Msun$, whereas the secondary mass is $1-4\Msun$. In Case AS mergers, mass transfer leads to a reversal of the mass ratio. As a result, the mass of the primary is approximately 2--4\Msun at the onset of merging and that of the secondary is 6--9\Msun (see also \autoref{fig:mergers_appendix} in the appendix) . 

In both cases, the stars come into a deep over-contact configuration \citep{Pols1994a, Wellstein+2001, de-Mink+2007}. In our simulations, contact results in merging of the stars as this is the most likely outcome \citep[see also discussion by][]{de-Mink+2014}.  
The simulations of  \citet{Gaburov+2008} and \citet{Glebbeek+2013} suggest that the merger product that forms will initially be a puffed up and red object \citep{Tylenda+2011,Pejcha+2016}, but as it radiates away the excess energy, it will recover a thermal equilibrium structure.  The core of the primary is thought to sink to the center of the new object with the core of the less evolved secondary settling above  \citep{Glebbeek+2013}.  The role of mixing and mass loss during this stage is highly uncertain. To first order, the further evolution may be approximated by that of a rejuvenated (rotating) single star.  It will continue with the subsequent burning stages and, if massive enough, end its life as its core collapses.  Whether or not the abnormal interior chemical profile or enhanced budget of angular momentum leaves significant imprints on the final pre-explosion structure and leads to observable signatures remains to be investigated.

\item[\bf HG$\Rightarrow$MS: ]

Forward mergers involving a post-main-sequence primary star are dominated by cases where the primary star is crossing the Hertzsprung gap (HG). These  so-called Case B mergers account for 15\% of the late ccSNe.  They originate from systems with unequal masses ($q<0.4$ in our standard simulations).  The systems come into contact as a result of the expansion of the secondary in response to rapid mass accretion and/or, for the most extreme mass ratios, because mass transfer is unstable \citep[see however][]{Pavlovskii+2016}. 
This leads to a common envelope phase that may lead to the ejection of (part of) the envelope while the orbit shrinks. If the ejection fails, we are left with a merger of a partially stripped post-main-sequence primary with its less evolved main-sequence secondary. The typical mass of the primary is 6--7\Msun at the onset of the merger. At this stage, it has an inert helium core of 1--1.4\Msun.  The typical secondary star has a mass of 1--3\Msun in our standard simulations (see also \autoref{fig:mergers_appendix} in the appendix). 

Simulations by \citet{Glebbeek+2013} indicate that the helium core of the original primary will sink to the center. The steep mean molecular weight gradient will likely inhibit mixing between the core and layers above. After the merger product has radiated away the excess energy  and restores thermal equilibrium, it has the structure of a giant with a core mass that is abnormal for its total mass compared to a regular giant created through single-star evolution.  A fraction of these objects are expected to appear as blue supergiants. The helium core grows as a result of shell hydrogen burning and, when massive enough, continues with the advanced burning stages.  The final pre-explosion properties are uncertain but some of these may give rise to 1987-like events \citep[e.g.,][]{Podsiadlowski+1990}; see also \citet{Vanbeveren+2013} and \citet{Justham+2014}.

\end{description}

\paragraph{Reverse mergers -- \label{sec:reverse_mergers}}

Approximately half of the late ccSNe (47\%) in our standard simulation (equivalent to 7.5\% of all ccSNe) originate from mergers following a reverse mass-transfer phase from the initially less massive secondary back to the initially more massive primary. Nearly all these systems experienced one or more semi-conservative phases of mass transfer from the primary to the secondary before, allowing the secondary to grow beyond the mass threshold  \Mmincc. The secondary would have exploded as a ccSN as a result of this.  However, as it evolves and expands, it encounters the stripped remnant core of the primary, which is still in orbit, triggering a reverse mass-transfer phase. The donor star is now  typically a few times more massive than the remaining core of the primary, leading to common envelope. The secondary engulfs the primary.  The orbit shrinks while part of the envelope of the secondary is ejected. If complete ejection is unsuccessful, the two stars will coalesce.
We distinguish different cases depending on the evolutionary stage of the primary and secondary.

\begin{description}

\item[\bf HeStar$\Leftarrow$HG:] 
In our standard simulation, 20\% of the late events are produced by the merger of the naked helium core of the initially more massive primary star and a post-main-sequence secondary after a common envelope phase. These are cases where the combined helium core mass after the merger exceeds the single star threshold for ccSN \citep[e.g.,][]{Nomoto1984,Habets1986,Nomoto1987} and can produce a metal core mass of at least $M_{\min, \rm metal}$.  

The typical progenitor system involves two intermediate-mass stars with similar initial masses (q>0.6) and an initial period shorter than approximately $10$ days. The typical mass of the naked He primary star at the moment of merging ranges from 0.4 to 1.3\Msun, while the secondary star ranges from 6 to 13\Msun and has developed a He core of 1.5 to 3\Msun (see also \autoref{fig:mergers_appendix} in the appendix).

\item[\bf COWD$\Leftarrow$HG:] 
We find that 10\% of late ccSNe result from the merger of a massive recently formed CO white dwarf, being the remnant of the primary, which is engulfed by a HG secondary star.  
The typical mass of the COWD is approximately $0.9$\Msun, being the remnant of a star with initial mass around 5-8\Msun, although there is a small contribution of progenitors with masses of  9-12\Msun that results in more massive white dwarfs (see also \autoref{fig:mergers_appendix} in the appendix).  The first group originates from wide systems with periods $ P>100$~days experiencing almost non-conservative mass transfer. The second group involves systems in tighter orbits,  $P\lesssim 30$~days, that evolved fairly conservatively, allowing the secondary to almost double its mass.   

Assessing the outcome of these mergers will require future dedicated simulations. \citet{Sabach+2014} have addressed the question of what is the result of these reverse common envelope channels involving a white dwarf. Here we speculate on some of the possible outcomes. We expect the white dwarf to spiral in and sink to the center of the merger product, where it will find itself surrounded by the former helium core of the secondary, that is, by 1--3\Msun of helium.  The structure of the star is a post-core-helium-burning star, with an abnormal ratio for the core and shell mass.  Nuclear burning will proceed in a hydrogen burning shell (if hydrogen is left after the spiral in phase) and a helium burning shell. The CO core will grow due to ashes of the helium burning shell. This may lead to off-center ignition of carbon.  Carbon ignition may occur in non-degenerate conditions and lead to the formation of an ONeMg core, which may eventually collapse as a result of electron capture \citep[e.g.,][]{Nomoto1984}. 

Alternatively, if carbon ignition occurs under degenerate conditions, it could, in principle, lead to either a deflagration, leaving a neutron star remnant, or a detonation of the entire core, similar to a type Ia thermonuclear explosion \citep{Nomoto+1991}. The main difference from normal Ia supernovae is that it would occur inside a hydrogen envelope.
In the case of the detonation of the core without leaving any remnant, it may be similar to the ``type 1.5'' supernova that \citet{Iben+1983} proposed, which is the detonation of the CO core of an Asymptotic Giant Branch (AGB) star that reaches a mass close to the Chandrasekhar limit. It has been suggested that this can also be the final outcome of single intermediate-mass stars in very low metallicity environments  \citep[e.g.,][]{Zijlstra2004,Lau+2008} where the wind mass loss is low enough to allow the formation of such a massive CO core. As we show here, binarity may allow similar structures even at higher metallicities. The theoretical ``type 1.5'' supernova mechanism is suggested as a possible explanation for the observed class of thermonuclear explosions that show interaction with circumstellar material, sometimes referred to as Ia-CSM supernovae \citep[e.g.,  SN2002ic,][]{Hamuy+2003}.

A third possibility that we cannot exclude at present is that the effects of mass loss and possible dredge-up prevent the star from growing a core massive enough to undergo core collapse. 

\item[\bf ONeWD$\Leftarrow$HG:]   Mergers involving an ONeMg white dwarf with and a HG secondary account for 11\% of the late ccSN in our standard simulations. The white dwarf has a mass between 1\Msun and the Chandrasekhar mass. The mass of the secondary is around 5--8\Msun at the moment of merging with a He core of around 1\Msun (see also \autoref{fig:mergers_appendix} in the appendix). These systems originate from initially wide systems (approximately 60-1000 days) that evolve very non-conservatively. We expect the products of these mergers to end their lives either as electron capture supernovae or iron-ccSNe (see also \citealt{Sabach+2014}).

\item[\bf COWD$\Leftarrow$AGB:] Mergers where the secondary is an AGB star merging with the CO white dwarf remnant of the primary contribute 6\% of the late events in our standard simulations.  In these cases, the combined mass of the white dwarf primary and the degenerate CO core of the secondary exceeds the Chandrasekhar mass. These systems primarily originate from progenitors with intermediate periods of 30--100 days (see also \autoref{fig:mergers_appendix} in the appendix).

One of the earliest discussions of this channel is given by \citet{Sparks+1974} who suggested that mergers of this type 
 may lead to collapse forming a neutron star.  This scenario differs from the classical case of the merger of two white dwarfs as the merger will occur inside a hydrogen envelope.   While this was originally discussed as a channel for thermonuclear explosions of white dwarfs \citep{Iben+1984,Webbink1984},  \citet{Nomoto+1985}, \citet{Saio+1985} and, more recently,  \citet{Schwab+2016}, argue in favor of the formation of a neutron star.

Regardless of whether this leads to a thermonuclear or core-collapse event, it is interesting to note that the timescale between the merger and the final explosive event is short. Remainders of the partially ejected envelope may still be in the close vicinity of the merger product, allowing for observational consequences as the supernova shock interacts with the circumstellar material.

Alternatively, \citet{Hachisu+1999} suggest that systems consisting of a COWD with a massive AGB star can avoid common envelope phase due to high wind mass-loss rates from the WD.  If the WD accretes enough mass to explode, it would provide an alternative possible channel for Ia-CSM supernovae  \citep{Nomoto+2004}.

\end{description}

\paragraph{Non-merger channels -- \label{sec:non_mergers}}

Approximately 13\% of the late ccSNe occur in binaries in which the stars interact but avoid coalescence. In nearly all cases, it is the initially less massive secondary that explodes, having previously accreted mass from the primary. In more than half of these cases, the primary explodes prior to the secondary, disrupting the system. The secondary explodes later after having drifted away from its place of birth by up to hundreds of parsecs.  

A few percent of the late ccSNe occur when a common envelope evolution initiated by the expansion of the secondary results in a successful envelope ejection. The outcome is a ccSN explosion from the stripped secondary next to an also stripped primary or a white dwarf. If the binary system does not get disrupted by this late supernova, it may lead to a system similar to PSR B2303+46 and PSR J1141-6545 in which the neutron star pulsar was formed after the WD companion \citep[e.g.,][]{Portegies-Zwart+1999a,Tauris+2000,Davies+2002,Kalogera+2005,Church+2006}.

\section{Robustness of the predictions} \label{sec:robustness}

To estimate the sensitivity of our results to variations in the assumed parameters, we conduct a set of simulations in which we vary the assumptions concerning the  physical processes (\autoref{subsec:physical_assumptions}) and the treatment of initial conditions (\autoref{subsec:initial_conditions}). In \autoref{table:parameters_uncertainties}, we provide a summary of the considered variations and the resulting fraction of late ccSNe $f_{\mathrm{late}}$.   The maximum range we find among all variations we consider is $f_{\mathrm {late}} = 7.2-24.3\%$.  
We also provide the median delay time of late ccSNe ($t_{\mathrm {50\%,late}}$) and the delay time below which 90\% the late events are found ($t_{\mathrm {90\%,late}}$). We discuss the most important effects below. 

\subsection{Physical processes}

The efficiency of mass transfer, that is, the fraction of the mass lost by the donor star that is accreted by the companion, is one of the main uncertainties affecting the evolution of binary systems \citep[for a discussion we refer to][]{de-Mink+2007}.  Our standard model considers a physically motivated simple prescription, which considers the thermal timescale at which the accreting star can respond to mass loss (\autoref{subsec:physical_assumptions}).  Lowering the accretion efficiency $\beta$ reduces the number of secondary stars that gain enough mass to exceed the threshold for ccSNe. This affects the reverse merger and the non-merger channels that produce late events. In the extreme case of fully non-conservative mass transfer ($\beta$ = 0, m01 in \autoref{table:parameters_uncertainties}), we still find that approximately $9\%$ of the ccSNe are late. Adopting  the opposite extreme of conservative mass transfer ($\beta$ = 1, m03) results in an increase of the fraction of late supernovae close to  $22\%$. 

Our results are also sensitive to the treatment of common envelope ejection \citep{Ivanova+2013}.  In our standard simulations, the large majority of late ccSN result from systems that merge during a common envelope phase.   Increasing the efficiency parameter for common envelope ejection  $\alpha_{\rm{CE}}$ enhances the fraction of systems that avoid coalescence.  Adopting a ten times larger efficiency  ($\alpha_{\rm{CE}} =10$, m17) prevents nearly all reverse mergers. Nevertheless, in this case, we still find a fraction of 7.2\% for the contribution of late ccSNe, from forward mergers or systems that do not merge. On the other hand, reducing the common envelope efficiency leads to an enhancement of $f_{\mathrm {late}}$ because of an increase in the number of merger products that produce a late ccSN. Simultaneously, this reduction of $\alpha_{\rm{CE}}$ augments the number of systems that merge prematurely during the first phase of mass transfer, hence resulting in a rearrangement of the relative importance of each evolutionary channel that produces a late event.

A few reverse channels for late events in our default simulation involve the merger of a white dwarf and a non-degenerate secondary. The outcome of such a merger is hard to predict reliably without further detailed simulations. To account for this uncertainty, we consider a model variation where we only allow for ccSNe from systems where the HG companion exceeds the threshold of $M_{\min, \rm cc}$ before merging, that is, where we would expect a ccSN even without merging with the white dwarf. The number of late events is reduced with this assumption, as expected, but the effect is mild: $f_{\mathrm {late}} \sim 13.2\%$ (m27). 

The formation of black holes in failed explosions may not be accompanied by a bright optical transient.  We therefore consider a model variation in which the most massive progenitors are excluded. Recent work suggests that there is no simple mass threshold for black hole formation \citep[e.g.,][]{OConnor+2011,Sukhbold+2016}, but this simple approach is sufficient to have an impression of the impact. We assume that all single stars with initial masses above a certain threshold , $M_{\max, \rm cc}$,  lead to no or very faint transients. For binary products, we consider the final metal core mass and compare with the resulting threshold for single stars.  Setting $M_{\max, \rm cc} = 35\Msun$ or $20\Msun$ gives  $f_{\mathrm late}= 16.9\%$ and  $20.4\%$ respectively (m28 and m29). The increase in the relative importance of the late ccSNe occurs because the excluded events have short delay times, thus the total number of ccSNe decreases but the tail is not affected.

The mass ratio criterion for contact and merging during the main sequence (m19-20) only affects the frequency of late ccSN by a few percent because it influences only one evolutionary channel. The criterion for unstable mass transfer when the donor is crossing the HG (m21-24) affects the results more, as many late events originate from systems that experience this process. Increasing $q_{\mathrm{crit,HG}}$, that is, allowing for unstable mass transfer in more systems even when the mass ratio of the two stars is not extreme, increases $f_{\rm {late}}$ to $\sim 22\%$ with the HG+MS forward merger becoming the dominant channel. 

The considered variations in  the minimum metal core mass for collapse ($M_{\min, \rm metal}$, m31 and m32) effectively change the minimum initial mass of a single star for core collapse, $M_{\min, \rm cc}$, to approximately 7 and $8\Msun$ respectively. Thus, the maximum delay time of single stars $\tau_{\max, \rm cc}$ also changes. We define, as `late' ccSNe, those that occur after the $\tau_{\max, \rm cc}$ for \emph{each} model. Thus, the median and 90th percentile of the delay time of late ccSNe is shifted, whereas the ratio of these events to the total is not significantly affected. 

We find that the frequency of late events is not sensitive to the adopted parameters for the treatment of main sequence mergers (m06-09). These mergers are not a dominant channel for late events.  Similarly, varying the assumptions concerning the supernova kicks, which can potentially disrupt the system (m10-11), has little effect because the majority of late events result from intermediate-mass systems in which only one SN occurs. Augmenting the angular momentum loss due to non-conservative mass transfer (m05) results in closer systems after mass transfer and thus to a slight increase in the number of mergers that lead to late ccSNe. Varying the stellar wind mass-loss efficiency (m25-26) does not significantly change $f_{\rm {late}}$.

\subsection{Initial conditions}

Of the adopted initial conditions, we find that the high-mass slope of the IMF, $\alpha$, is the largest source of uncertainty, in agreement with the conclusions drawn by \citet{de-Mink+2015}.  A steeper IMF favors late supernovae as they originate from intermediate-mass progenitors. We find variations of $f_{\rm {late}} = 8.2-24.3\%$  when varying the slope from $\alpha = -1.6$ to $-3$ (m32-34), which are the extreme values considered. 

Late events preferentially originate from systems with comparable initial masses  and initial periods less than approximately 30 days (see \autoref{fig:progenitors}).   Adopting an initial mass ratio distribution that more strongly favors equal mass systems thus leads to a relative increase of late supernovae. We obtain $f_{\rm {late}} = 12.1-18.1\%$ when varying the slope of $\kappa$ from $-1$ to $+1$ (m35-36). Similarly, a period distribution favoring short period systems also enhances the relative fraction of late supernovae. We find $f_{\rm {late}} = 11.3-23.2\%$ when varying $\pi$ from $-1$ to $+1$ (m37-38). 

Changing the overall binary fraction, $f_{\rm bin}$, proportionally affects the contribution of late supernovae, with $f_{\rm {late}} = 7.7-20.0\%$ when varying $f_{\rm bin} = 0.3, 1.0$ (m46-47).  
In addition, we also compute a population assuming a mass dependency in the binary fraction, which favors binarity in more massive stars (m48, \autoref{eq:fbin_mass}).  This results in a decrease of $f_{\rm {late}}$ to  $12.8\%$, as intermediate-mass binaries that are the progenitors of late ccSNe are disfavored. We have tested that this result is not very sensitive to the actual choice of the mass boundaries in \autoref{eq:fbin_mass}. The normalization parameter, $M_{\mathrm{low}}$, which affects the total mass contained in low-mass stars, does not affect the shape of the DTD, only the total number of ccSNe per unit mass (m48-49).

We also consider the effect of metallicity. The minimum mass for core collapse in a single star ($M_{\min, \rm cc}$) decreases for lower  metallicity \citep{Pols+1998} and thus the lifetime of the last ccSN resulting from single stars $\tau_{\max, \rm {cc}}$ increases. For example, for $Z=0.002,$ we find $M_{\min, \rm cc} = 6\Msun$, which corresponds to a lifetime of approximately $70$ Myrs.  In general, we thus find a shift to longer delay times for lower metallicities. 
We find that $f_{\rm {late}}$, defined as the fraction of  supernovae that explode after the last single supernova for \emph{that} metallicity, decreases.  This is because stars at low metallicity are generally more compact. In a binary system with the same initial parameters, that is, $M_1$, $M_2$, and period, mass transfer typically occurs at a later evolutionary stage of the primary in a lower metallicity system. This disfavors most of the evolutionary channels that produce late events.  

\begin{table}[th!]
  \caption{Comparison of the median delay time of all core-collapse supernovae, $t_{\mathrm {50\%, all}}$, and the delay time below which 90\% of all events are found, $t_{\mathrm {90\%, all}}$, in simulations by independent groups. The fraction of late ccSNe $f_{\rm late}$ is also shown. All simulations including binaries adopt a binary fraction of 0.7. \label{table:comparison_median_90}}
  \centering
   {\small
  \begin{tabular}{ll ccc }
 \hline 
\hline
\multicolumn{2}{l}{Simulation} &   $t_{\mathrm {50\%, all}} $ & $t_{\mathrm {90\%, all}} $   & $f_{\rm late}$ \\
& &   (\Myr)  & (\Myr) & \\
\hline
\multicolumn{4}{l}{\it Single stars} & \\
- &STARBURST99 (default)  &             15.0 &            32.7 & 0\% \\
- &STARBURST99 (modified) &             16.4 &            36.7  & 0\%\\
- &BRUSSELS  &             23.1 &            39.7  &0\%\\
- &BPASS  &             17.0 &            50.1 &0\% \\
- &{\bf this work }&              {\bf 19.6 } &          {\bf   41.6 } & {\bf 0\%} \\
~\\
\multicolumn{4}{l}{\it Including binary stars} & \\
- &BRUSSELS    &            31.5 &           112.7 &  23\%\\
- &BPASS  &             17.7 &            63.1 & { 8.5\%} \\
- &{\bf this work }&            {\bf  22.1 } &       {\bf      64.1 } & {\bf 15.5\%}\\
\hline
\hline
  \end{tabular}
  }
\end{table}

\subsection{Comparison with results by independent groups\label{sec:Comparison}}

Various independent groups provided predictions for the DTD of ccSNe.  In \autoref{table:comparison_median_90}, we show the results of different groups for the fraction of late ccSNe, $f_{\rm late}$, as well as the median delay time,  $t_{\mathrm {50\%, all}} $, and the delay time before which 90\% of all ccSNe are found, $t_{\mathrm {90\%, all}}$.  We find that $f_{\rm late}$ predicted by different groups that consider binarity agrees within a factor of two with our standard predictions. Simulations that only account for single stars find $f_{\rm late} = 0\%$, by definition.   Below, we discuss the individual studies.

\paragraph{STARBURST99}
-- We use the online interface  of STARBURST99 \citep{Leitherer+1999,Vazquez+2005,Leitherer+2010,Leitherer+2014} to compute the DTD, using two sets of assumptions. In the first, labelled  ``default'', we choose the default assumptions of the online interface that use the classical grid of models by \citet{Schaller+1992} for $Z=0.02$, adopting a \citet{Kroupa2001} IMF and  assume $M_{\min, \rm cc} = 8\Msun$. We also compare our results with predictions obtained with ``modified'' assumptions where we set the threshold to $M_{\min, \rm cc} = 7.53\Msun$ and $Z = 0.014$, to match our standard assumptions, and where we selected the newer non-rotating models by \citet{Ekstrom+2012}.

\begin{figure*}
\centering
\includegraphics[width=0.7\textwidth]{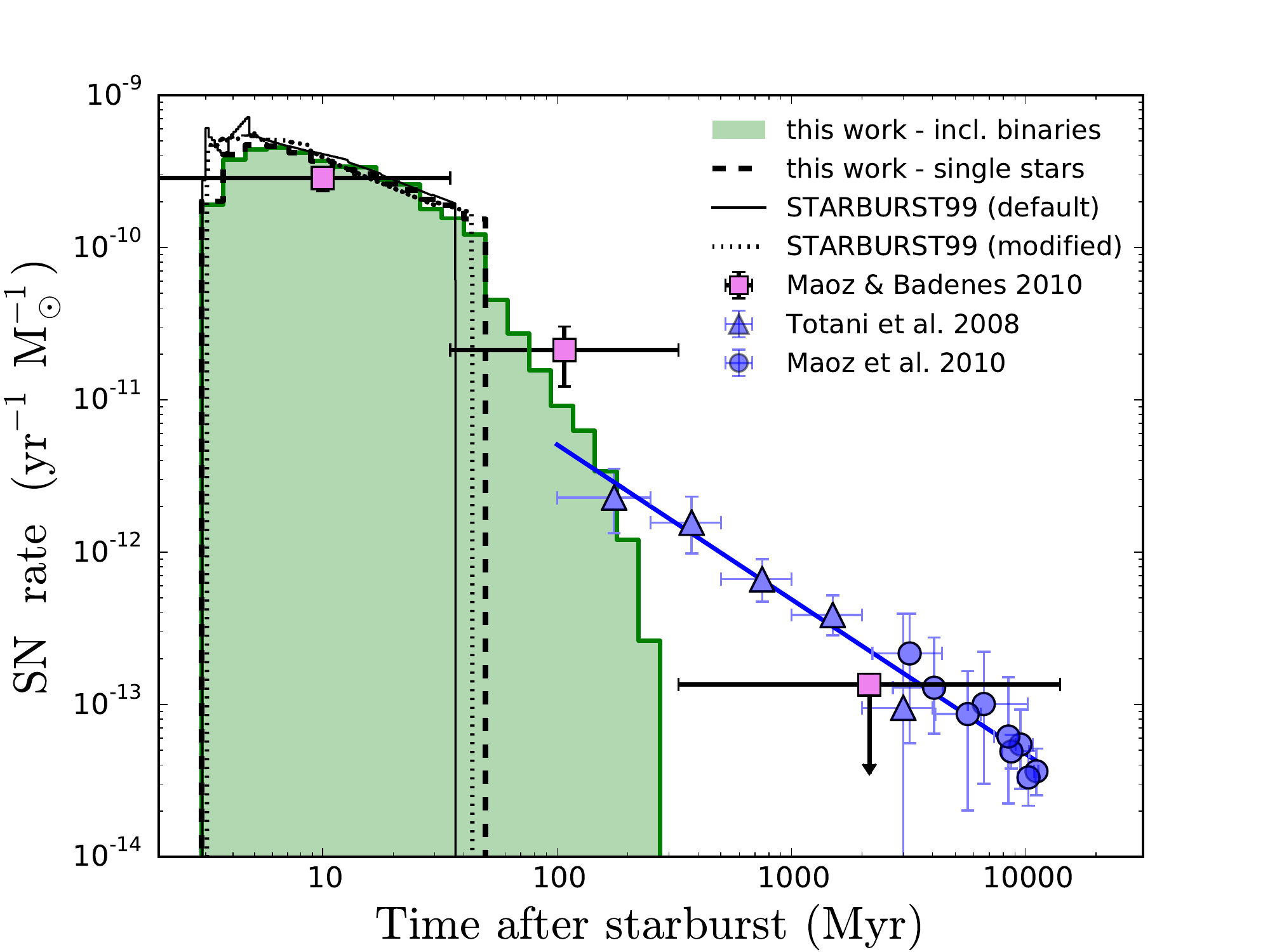}
\caption{ 
Comparison of our delay-time distribution of ccSNe including binaries (green shading)  with those of our single-star population (black dashed line) and predictions by STARBURST99 (thin full and dotted lines, see \autoref{sec:Comparison} for a description). We further show measurements by \citet{Maoz+2010} using Magellanic Cloud supernova remnants (pink squares), which contain a combination of remnants of ccSNe and SNe Ia.  Blue triangles and circles represent the observed rates of SNe Ia measured in extragalactic surveys in \citet{Totani+2008} and \citet{Maoz+2010a}, respectively.  The blue line is a $t^{-1}$ fit through the extragalactic data. Horizontal error bars indicate the bin size. Vertical errorbars show the $1 \sigma$ error bars on the rate. The rightmost pink supernova rate is an upper limit.
We argue that `late' ccSNe can help explain the discrepancy between the intermediate age bin in the DTD by \citet[][central pink square]{Maoz+2010} and the trend in the extragalactic SNe Ia (blue line), instead of the original interpretation that this age bin includes only `prompt' SNe Ia (see discussion in \autoref{test2}).
}
\label{fig:DTD_ia_badenes_data}
\end{figure*}

The DTD of the STARBURST99 simulations is shown in \autoref{fig:DTD_ia_badenes_data} along with our standard simulations of single stars and including binaries (we note that the units on the y-axis are different than in Figures \ref{fig:SN_delay_time_single} and \ref{fig:SN_delay_time}; for a discussion, we refer to \autoref{Appendix:fit}). The DTDs for single stars are very similar until the maximum delay time, which is approximately 35\Myr in the default STARBURST99 population, 40\Myr for the modified STARBURST99, and 50\Myr in our simulation with only single stars. This is due to differences in the minimum single star threshold mass for ccSN as well as differences in lifetimes (see also \autoref{fig:lifetime}). 

When we account for binaries, we find an increase of more than 30\% in the median of the DTD ($t_{\mathrm {50\%, all}} = 21.5 \Myr$, \autoref{table:comparison_median_90}) with respect to the default STARBURST99 models ($\sim 16$ \Myr). The 90th-percentile of the distribution, $t_{\mathrm {90\%, all}} $, shifts by almost a factor of two (from 33 \Myr to 64 \Myr) because of the tail of late events that are produced only by binaries.

\paragraph{BRUSSELS Group}
 -- \citet[][henceforth DV03]{De-Donder+2003} were the first to explore the effects of binaries on the DTD of ccSNe and find very similar general trends. They find a tail of late events after their last single star explodes at approximately 40\Myr extending until approximately 250\Myr, slightly later than what we find (200\Myr). 
They describe a contribution from mergers where the secondary engulfs the white dwarf remnant of the primary, which is responsible for $\sim 16$\% of all ccSNe in their simulations. This is significantly higher than what we find ($\sim 4$\% of all ccSNe, combining channels of CO/ONeWD$\Leftarrow$HG/AGB in \autoref{table:channels}). They mention that the secondary is either still on the main sequence or is an evolved red giant with a CO core.  In our simulations, we find instead that most of these mergers result from cases where the secondary is a HG star. The authors do not mention channels where the remnant of the primary is a naked helium star, which is the dominant reverse channel in our simulation (HeStar$\Leftarrow$HG in \autoref{table:channels}).
 
The original data from their simulations is no longer available (D. Vanbeveren, priv. communication, 2015), but we extract the data from the top panel of Figure~1 in \citet{De-Donder+2003} using the PlotDigitizer software tool distributed under the terms of the GNU General Public License. For a binary fraction of 0.7, they find a total fraction of late events of approximately 23\%, which is higher than our 15.5\%. In part, this is due to the steeper IMF that these authors adopt, $\alpha = -2.7$.  Our model 33, which adopts the same IMF as DVO3, results in a fraction of late events of 20.4\%.  

The BRUSSELS simulations show significantly longer delay times. They find a median delay time of 32\Myr, which is approximately 50\% larger than our standard simulation. Their 90th-percentile lies at 113 \Myr,   which is nearly twice as large as in our standard simulations; see \autoref{table:comparison_median_90}.

\paragraph{BPASS  }
-- \citet{Xiao+2015} discuss predictions for the ccSNe rate using the  Binary Population and Spectral Synthesis (BPASS) code by \citet{Eldridge+2009}.  They find a maximum delay time for core-collapse events from single stars around 70\Myr, which is considerably larger than in our simulations ($\sim 50$\Myr).  No tail of late events due to binary interactions is visible in their Figure~1, maybe because these simulations do not include intermediate-mass binaries. 

The newer BPASS v2.0 models are based on more extended model grids. We retrieved them from the online repository. The coverage of the initial binary parameter space is still sparse for our purposes ($\sim$ 5000 binary systems per metallicity, while we use approximately two orders of magnitude more systems in our variations and three orders of magnitude more for our standard simulation, see \autoref{subsec:simulation_setup}). However, we find similar trends. The BPASS v2.0 models also show a tail of late events up to approximately 250\Myr after starburst, comparable to our findings.  The late events in the BPASS models account for 8.5\% of all ccSNe, which is lower than but still consistent with our findings. The median and 90th-percentile of their distribution are similar to ours, see  \autoref{table:comparison_median_90}.  

\section{Discussion} \label{sec:Discussion}

We discuss a variety of ideas to observationally test the predicted tail of late ccSNe and highlight current observations that already provide hints (section \ref{test1}-\ref{test3}).  We then discuss possible implications for stellar feedback and for the derived cosmic-star-formation rate. 

\subsection{Direct progenitors and descendants of late ccSNe  \label{test1} }

Our simulations predict the existence of direct progenitors of late ccSNe. The well studied, post-interaction binary $\phi$ Persei \citep[][]{Poeckert1981,Slettebak1982} provides  a very strong case of such. 

The brightest star in the system is a Be star with a mass of ${9.6 \pm 0.3\Msun}$ \citep[][and references therein]{Mourard+2015}; well above \Mmincc.
This binary system is considered to be a member of the  $\alpha$~Persei cluster, which has an estimated age of 50--60\Myr  \citep{Makarov2006,Zuckerman+2012}. This is almost twice the expected lifetime of a single star with a mass similar to that of the Be star.  This age discrepancy can be naturally explained if the Be star was born with lower mass before gaining mass from a companion later in its life, giving it the appearance of a blue straggler.  The rapid rotation of the Be star is a further indication that it was spun up during a previous phase of mass transfer \citep{Packet1981, de-Mink+2013}. 

Evidence is also provided by the properties of the companion, which is a hot subdwarf O star of $1.2\pm0.2\Msun$ \citep{Mourard+2015}. Its  properties and current orbital period of 126 days match the theoretical expectations for a typical post-mass-transfer system. The hot, low-mass companion is the remaining stripped helium core of the initially most massive star that lost its hydrogen-rich envelope during Roche-lobe overflow.  The current mass of the subdwarf corresponds to the core mass of a star with an initial mass of approximately 7\Msun.  Evolutionary modeling of the system by different groups show that the Be star was likely born with a mass around 4--5\Msun  \citep[][Schootemeijer et al. 2017, in prep.]{Vanbeveren+1998, Pols2007}. 

This means that both stars were born with masses below the single-star mass threshold for ccSN.  Binary interaction increased the mass of the Be star sufficiently that is should end its life as a ccSN. However, the final fate of the system depends on the uncertain phase of common envelope evolution that is expected when the Be star leaves the main sequence, expands and engulfs the subdwarf.  If the system fails to eject the envelope, which we deem likely, the resulting merger product will likely be massive enough to undergo core collapse. If the system successfully ejects the envelope, the outcome will depend on how much mass is lost during subsequent mass transfer phases. 

More generally, as mentioned in \autoref{sec:reverse_mergers}, if a similar system survives both the reverse common envelope process and the ccSN of the secondary, it may lead to an eccentric binary of a white dwarf and a neutron star, similar to PSR B2303+46 and PSR J1141-6545. It has been argued that in these systems, the secondary formed a neutron star after the primary became a white dwarf \citep[e.g.,][]{Portegies-Zwart+1999a,Tauris+2000,Davies+2002,Kalogera+2005,Church+2006}, meaning the neutron star was formed during a `late' ccSN. For example, \citet{Tauris+2000} discuss such an evolutionary scenario for PSR J1141-6545, predicting that the supernova occurs at approximately 70 \Myr . This would imply that PSR B2303+46 and PSR J1141-6545 systems are direct descendants of `late' ccSNe.

Progenitor systems such as $\phi$ Persei are extremely valuable for verifying the predictions of the evolutionary models. FY Canis Majoris \citep{Peters+2008} and 59 Cygni \citep{Peters+2013} are systems that probably followed a similar evolution with $\phi$ Persei and may also be progenitors of late ccSNe.
Only a small number of post-interaction systems have been identified and characterized so far, due to the many biases that hinder the detection of these systems \citep{de-Mink+2014}.  To move forward, it is necessary to prioritize systematic observing campaigns that are carefully designed to identify them (e.g., G\"otberg et al. 2017, in prep.) to provide more stringent test cases with well determined parameters. 

\subsection{A possible signature of late ccSNe in the sample of Magellanic Cloud supernova remnants
 \label{test2}}

Several groups have tried to measure the DTD of supernovae. However, in general, all studies focus on the delay time of type Ia supernovae (SNe Ia), which are thought to arise from the thermonuclear explosion of a white dwarf  \citep[e.g.,][]{Totani+2008,Maoz+2010a,Maoz+2011,Maoz+2012,Maoz+2010,Graur+2013,Graur+2014}; we refer to \citet{Maoz+2014} for a review. The general idea behind the different techniques that are used is to measure SN rates in different environments and compare them to the star-formation history (SFH) of the associated stellar populations (e.g., the cosmic SFH based on the redshift of the explosion or the SFH of each galaxy for extragalactic surveys, etc.). 

Because SNe Ia originate from lower-mass systems than ccSNe, they typically have longer delay times; of the order of hundreds of Myr to Gyr, in contrast to the typical tens of Myr for ccSNe. This can be seen in \autoref{fig:DTD_ia_badenes_data} where we show the DTD of SNe Ia derived by \citet{Totani+2008} and \citet{Maoz+2010a} from extragalactic surveys (blue triangles and circles, respectively). The blue line is a simple power-law fit to highlight the widely accepted $t^{-1}$ form of the SNe Ia DTD \citep[e.g.,][]{Maoz+2014}.

The techniques used to measure the DTD of SNe Ia cannot be applied directly for ccSNe. The much shorter typical delay times of ccSNe require a much higher accuracy in the derived age of the associated stellar populations, for any meaningful comparison of data and models. Even in \citet{Maoz+2011}, where the time accuracy is better than all previous methods, all ccSNe are grouped together in the first time bin ($0-420$ \Myr).  Thus, it is difficult to observationally probe the short time-delay regime where the model predictions of ccSNe DTD lie.

\citet{Maoz+2010} pioneered a new technique to statistically infer the DTD, based on the largely complete sample of the 77 supernova remnants in the Magellanic Clouds \citep{Badenes+2010}. Because of the proximity to the Clouds, the stellar populations in the vicinity of the remnants can be resolved. This allows for a more accurate estimate of the age distribution of the surrounding population, providing statistical constraints on the delay time.  We show their results in \autoref{fig:DTD_ia_badenes_data} (pink squares). This is the highest-quality DTD at early times, probing the short delay regime and so far the only published DTD based on resolved stellar populations. 

While the original main interest of these authors was to probe SNe Ia with short delay times, their sample also includes remnants produced by ccSNe.  Unfortunately, for most of the remnants in this sample, it is no longer possible to reliably tell whether they came from type Ia or core-collapse supernova.  Most of them are more than 10~kyrs old and they are in the Sedov-Taylor phase \citep{Maoz+2010}. Therefore, the DTD inferred from this sample contains the combined signal of SNe Ia and ccSNe. 

\citet{Maoz+2010}  argue that the signal in the intermediate age bin that spans between 35  and 330\Myr (central pink square in \autoref{fig:DTD_ia_badenes_data}) is dominated by so-called prompt SNe Ia. They justify their reasoning with single-star models, which predict the last single star to explode at approximately 35-50 \Myr (\autoref{fig:lifetime}).  The authors also explicitly test their assertion by separately analyzing the subset of remnants for which the type of explosion is known.  Based on these, the authors conclude that `leakage' of core-collapse events from the first into the intermediate age bin is likely to be small. Unfortunately, the subset of remnants that could be used for this test is small because only 11 remnants have been identified as resulting from ccSNe.

Our simulations challenge the interpretation that the intermediate age time bin (35--330\Myr, central pink square) is purely due to prompt SNe Ia. Instead, our predictions suggest a significant contribution to this time bin from late ccSNe resulting from binary systems.  This can be seen in \autoref{fig:DTD_ia_badenes_data} where we overplot our model predictions for the DTD of ccSNe (green shading).  For comparison, we also show our predictions for single stars (black dashed line) and the predictions by the widely-used STARBURST99 population synthesis code for single stars  (thin full and dotted lines, discussed separately in \autoref{sec:Comparison}). 

Support for a significant contribution of late ccSNe comes from the discrepancy between the measurements by \citet[][]{Maoz+2010} for the intermediate age time bin (central pink square) and the general trend in the extragalactic SNe Ia (blue line). For these extragalactic measurements, there is no reason to confuse type Ia and core-collapse events. They are classified directly based on their light curves and spectra of the explosions.  We therefore consider the extragalactic measurements to be a reliable indication of the general trend in the DTD for SNe Ia.  These measurements suggest a much lower contribution of prompt SNe Ia to the intermediate age bin than \citet{Maoz+2010} (the offset between the blue line and central pink square).

We conclude that the data is consistent with a substantial contribution of late ccSNe. We refrain from drawing a firm conclusion, given the challenging nature of the measurements by \citet[][]{Maoz+2010} and the possibility that systematic errors play a role. However, the comparison above is certainly promising and shows the large potential of this method.  

\paragraph{Possible future improvements --}
A significant step forward can be made by more accurately characterizing the surrounding stellar population for each supernova remnant. The original study of \citet{Maoz+2010} was limited by the spatial and time resolution of the SFH map of the Magellanic Clouds by \citet{Harris+2004,Harris+2009}. \citet{Jennings+2012, Jennings+2014} and  \citet{Williams+2014} demonstrate how more accurate results can be obtained using archival Hubble Space Telescope imaging. This allows extension to other nearby galaxies, such as M31 and M33 \citep[see also][]{Murphy+2011}. 

Further improvement can be made with respect to the stellar models adopted for the analysis. As we show in \autoref{fig:lifetime}, more modern calibrated stellar models lead to lifetimes that are approximately 15\% longer.  Also, for consistency, the inclusion of the effects of binarity in the characterization of the stellar population is highly desired. This is now becoming possible as demonstrated by \citet{Wofford+2016}, using the BPASS simulations. This is important since products of binary evolution are typically rejuvenated and make the stellar populations appear younger than they actually are \citep[e.g.,][]{Schneider+2014}. Systematically underestimating the age of stellar populations surrounding the remnants will obstruct the identification of remnants resulting from late ccSNe. 

Progress can also be made by improving our understanding of the observational biases in the sample of remnants \citep{Sarbadhicary+2016} and by increasing the fraction of remnants where the type of explosion can be unambiguously determined.  \citet[][]{Lopez+2011} propose that the morphology of the remnants can be used to determine this.  X-ray spectral properties can also be used to classify a large fraction of the remnants \citep{Maggi+2016}.  A potentially promising strategy would be to target remnants surrounded by intermediate age populations for deep follow up and search for a remaining neutron star \citep[e.g.,][]{Perna+2008}. This is now also possible in the Magellanic Clouds, such as  PSR J0537-6910 in the remnant N157b, \citep[][]{Marshall+1998} for example.

Several studies focus on trying to identify runaway stars inside supernova remnants \citep{Tetzlaff+2013}. Our simulations imply that this is not a promising path to search for late ccSNe because the majority arise from mergers. However, this prediction is subject to uncertainties, most notably the poorly constrained common envelope ejection efficiency, $\alpha_{\rm CE}$. 

\subsection{ Photometric transient surveys: Possible expected signatures in the light curves of late events
\label{test3}}

If indeed approximately one in seven ccSNe are late, as is the case in our standard simulation, many of these events should already have been detected in supernova surveys. We expect the majority of events to be hydrogen rich type II supernovae, although the precise fraction is dependent on model assumptions such as the common envelope efficiency, $\alpha_{\rm{CE}}$. We note that the late events come from a variety of evolutionary scenarios (\autoref{table:channels}). Thus, they may not form a very homogeneous SN class and different distinguishing features may be expected among the different subtypes. Dedicated simulations will be required to predict the variety.  Here, we briefly speculate on variations that may qualitatively be expected. 

A subset of channels consist of mergers involving at least one star that has evolved beyond helium burning (i.e., the reverse mergers CO/ONeWD$\Leftarrow$HG/AGB, see \autoref{table:channels}, \autoref{sec:reverse_mergers}). In these cases, the supernova is expected to occur relatively quickly following the merger event. The merging is expected to lead to the shedding of several solar masses of hydrogen.  If a small fraction of this material remains in the vicinity until the explosion, the event may show signatures of interaction with a dense circumstellar medium and give rise to narrow emission lines (Type IIn).

We may expect further signatures in the light curve that indicate an abnormal pre-explosion structure resulting from mixing or mass loss during the binary interaction event.  For example, excessive mass loss may result in type II events with lower than usual mass in the hydrogen envelope. Another possibility could be that the envelope is more helium rich than usual, which may leave an imprint on the shape of the plateau of the light curve.

Finally,  the merger events themselves may constitute a subset of the rich and diverse class of the observed red intermediate luminosity transients or of gap transients, as has been pointed out by \citet{Ivanova+2013a}.

\subsection{Binarity and the missing supernova problem in relation to the cosmic star-formation rate}

Core-collapse supernovae can be used as tracers of the cosmic star-formation rate \citep[e.g.,][]{Horiuchi+2011,Madau+2014}.  This method makes use of the fact that the delay times between star formation and the resulting ccSNe are short compared to cosmological timescales.  This is still true when considering late ccSNe due to binarity because the impact on the average delay times is too small to have an effect.  Binarity does affect the total number of ccSNe produced per unit mass formed in stars (\autoref{totalNumber}).   \citet{Horiuchi+2011} pointed out that there is a discrepancy between the cosmic ccSN rate and the measured cosmic massive-star-formation rate, with the former being approximately a factor 2 lower than the latter \citep[cf.][]{Hopkins+Beacom2006,Mannucci+2007,Botticella+2008}. 
 
To arrive at this conclusion , \citet{Horiuchi+2011} use a conversion factor of  0.0088 ccSNe per \Msun to convert the cosmic star-formation rate into the expected core-collapse rate.  In comparison, we find a conversion factor of 0.0115 ccSNe per \Msun in our standard single-star simulation with large variations when changing the IMF (0.0028--0.0253 ccSNe per \Msun). In other words, uncertainties in the IMF alone can, in principle, already resolve the discrepancy.   Accounting for binarity increases this conversion factor by $14^{+15}_{-14}$\% to 0.0128 SNe per \Msun in our standard binary simulations  (0.0035--0.0253 SNe per \Msun in the variations considered), making the missing supernova problem a factor of 1.5 times worse. 

Binarity will also affect the derived star-formation rate itself as it changes the production of UV and H{$\alpha$} photons per unit mass, which are used as indicators. For a discussion of these effects we refer to \citet{Xiao+2015}  and references therein.

\subsection{Late core-collapse supernovae and implications for feedback}

Supernovae play a crucial role as sources of feedback in galaxies. Binarity is likely to affect when, where, and how this process occurs. The level of sophistication used to account for supernova feedback in simulations of galactic evolution varies between studies (if implemented at all). In one of the more  simple forms, all core-collapse explosions are assumed to happen promptly or at a fixed time after starburst, typically 10\Myrs.  We would advise using 20\Myrs, which corresponds to the median delay time ($t_{\mathrm {50\%, all}}$, \autoref{table:comparison_median_90}), as a more realistic value. Accounting for binary systems has little effect on the median (from 19.6\Myr for single stars to 22.1\Myr for simulations with binaries).  For a more realistic implementation, it is worth considering the spread in delay times. As mentioned in \autoref{sec:Comparison}, the delay time up to which 90\% of ccSNe have exploded, $t_{\mathrm {90\%, all}} $,  increases from 33\Myr in STARBURST99 to 42 \Myr in our standard single-star simulation and to 64\Myr when we include binary systems (\autoref{table:comparison_median_90}).

These differences are large enough to be worthy of consideration.  \citet{Kimm+2015} showed that using more realistic stellar lifetimes (meaning the extended delay times predicted by the single stellar models in the STARBURST99 simulations) has a significant impact on galactic models. It can for example prevent the formation of high density gas clouds and lead to some explosions occurring in low-density environments.  Binarity can stretch the DTD significantly.  \citet{Struck-Marcell+1987,Parravano1996,Quillen+2008} show that a delay between starburst and feedback is needed for episodic star formation to occur. In their estimates, \citet{Quillen+2008} assume ccSNe delay times of a few Myrs. If average supernova delay times are longer, as we show, the potentially increased feedback timescale may impact future star formation in the molecular cloud.

The delays also imply that the supernovae will occur significantly displaced from their birth location. A star with a delay time of 60\Myr that has inherited a velocity of approximately 5\kms from the turbulent motions of its birth cloud will travel 300 pc before exploding.  Note that this is without invoking any runaway nature.

Finally, a delay time implies a lag in chemical enrichment of the environment. It is unclear, at present, whether or not the tail in the DTD discussed here is significant enough to cause observable effects to chemical enrichment models.

\section{Summary}\label{sec:summary}

We investigate the impact of binarity on the stars that end their lives as core-collapse supernovae (ccSNe) and on the delay time between the starburst and their explosions. We use a binary population synthesis code to compute the delay-time distribution (DTD) of ccSNe, where we calculate the supernova rate versus time after a starburst. Our main findings are as follows:
\begin{itemize}
\item Binaries extend the DTD to longer delay times (50-200 Myr) compared to a population consisting only of single stars. The fraction of `late' ($>50$ Myr) to the total number of  ccSNe in a population with binaries is $f_{\rm late} = 15.5^{+8.8}_{-8.3}$\%.

\item  Late ccSNe originate predominantly from intermediate-mass binaries, in which one or, in three out of four cases, both stars have birth masses below $M_{\min, \rm cc}$,  the threshold for single stars to explode. The typical initial primary masses are between 5 and 8 \Msun.

\item With our standard assumptions, almost half of the late supernovae originate from merger events of a post-main-sequence star with either a stripped helium star or a white dwarf, following a common envelope phase. Prior to the merger, the system experiences one or more mass transfer episodes from the primary, that is, the initially more massive star that later becomes the helium star or the white dwarf, to its companion (secondary) that later evolves to the post-main-sequence star that initiates the merger. Late supernovae from this type of `reverse' mergers constitute approximately 7.5\% of the total number of ccSNe in our standard simulation. The slow evolutionary timescales of their progenitors, compared to more massive stars, are why most of them end up in the late tail of the DTD. In our code, these reverse mergers may lead to ccSNe but detailed models should further investigate the outcome of these intriguing products and thus we provide the mass distributions of the stars at the onset of merging.

\item Another significant channel for late supernovae is the merger of an intermediate-mass star with its companion during the first mass-transfer episode (`forward' mergers), contributing to approximately 25\% of all late ccSNe. The primary is near or immediately after the end of its main sequence life and its companion is usually fairly unevolved. The merger product is massive enough to experience collapse and is also rejuvenated, leading to a late ccSN. We provide the properties of the stars at the moment of merger for these channels as well. 

\item A smaller contribution to late events comes from systems that do not experience merging, either because of successful ejection of the envelope during a common envelope evolution phase or due to disruption of the binary system by a prior supernova explosion.

\item We test our results by varying the parameters of our standard model one by one. Our results are sensitive to only a few of them, mainly the mass transfer efficiency, the common envelope evolution parameter and the IMF slope. The fraction of late supernovae also decreases for lower binary fractions especially for stars in the intermediate-mass regime, because these binary systems are the main progenitors of late events. The error ranges provided in all our results are based on these variations.

\item We find that populations including binaries produce $14^{+15}_{-14}$\% more ccSNe than singles stars, for the same amount of stellar mass.

\item We discuss various possible ways to observational test our predictions. We argue that the binary system $\phi$ Persei is a direct progenitor system and that the Be star in this system will likely give rise to a late ccSN. We further argue that the eccentric neutron star -- white dwarf binaries, in which the neutron star is believed to have formed after the white dwarf,  are direct descendants of late ccSNe, in agreement with earlier studies.

\item We discuss how age dating the stellar populations at supernova explosion sites and around supernova remnants can constrain the DTD of ccSNe.  In particular, we argue that the excess of Magellanic Cloud supernova remnants surrounded by intermediate age populations (spanning 35--330\Myr), that \citet{Maoz+2010} provide, can be interpreted as evidence favoring the existence of late ccSNe (effectively reducing the contribution of prompt Ia supernovae).

\item We briefly discuss the potential implications for the cosmic supernova rates and role of late ccSNe as sources of feedback.  We provide fitting formula in the Appendix to facilitate a simple implementation of our results in simulations.

\end{itemize}

\begin{acknowledgements}
 We want to thank the anonymous referee for the very useful and constructive comments. We are grateful for enlightening discussions with Nathan Grin, Rubina Kotak, Laura Lopez, Tom Maccarone, Maryam Modjaz, Maxwell Moe, Ken Nomoto, Onno Pols and Fabian Schneider. EZ is supported by the Netherlands Research School for Astronomy (NOVA). SdM acknowledges support by a Marie Sklodowska-Curie Action Incoming Fellowship (H2020 MSCA-IF-2014, project id 661502). RGI thanks STFC for his Rutherford fellowship (ST/L003910/1), the DAAD for funding TS, and Churchill college for funding his bi-fellowship and for access to their library. CB acknowledges NASA ADAP grant NNX15AM03G S01 and NSF/AST-1412980. SCY was supported by the Korea Astronomy and Space Science Institute under the R\&D program (Project No. 3348-20160002) supervised by the Ministry of Science, ICT and Future Planning. The authors further acknowledge the Leiden Lorentz Center workshop ``The Impact of Massive Binaries Throughout the Universe'' and the Munich Institute for Astro- and Particle Physics (MIAPP) of the DFG cluster of excellence ``Origin and Structure of the Universe'' for supporting the ``Physics of Supernovae''. We thank Thomas Russell for help in language editing.

\end{acknowledgements}

\bibliography{my_bib,manos_additions}

\begin{appendix}

\section{Supernova rate and fitting formula \label{Appendix:fit}}
\begin{figure}
\centering
\includegraphics[width=0.5\textwidth]{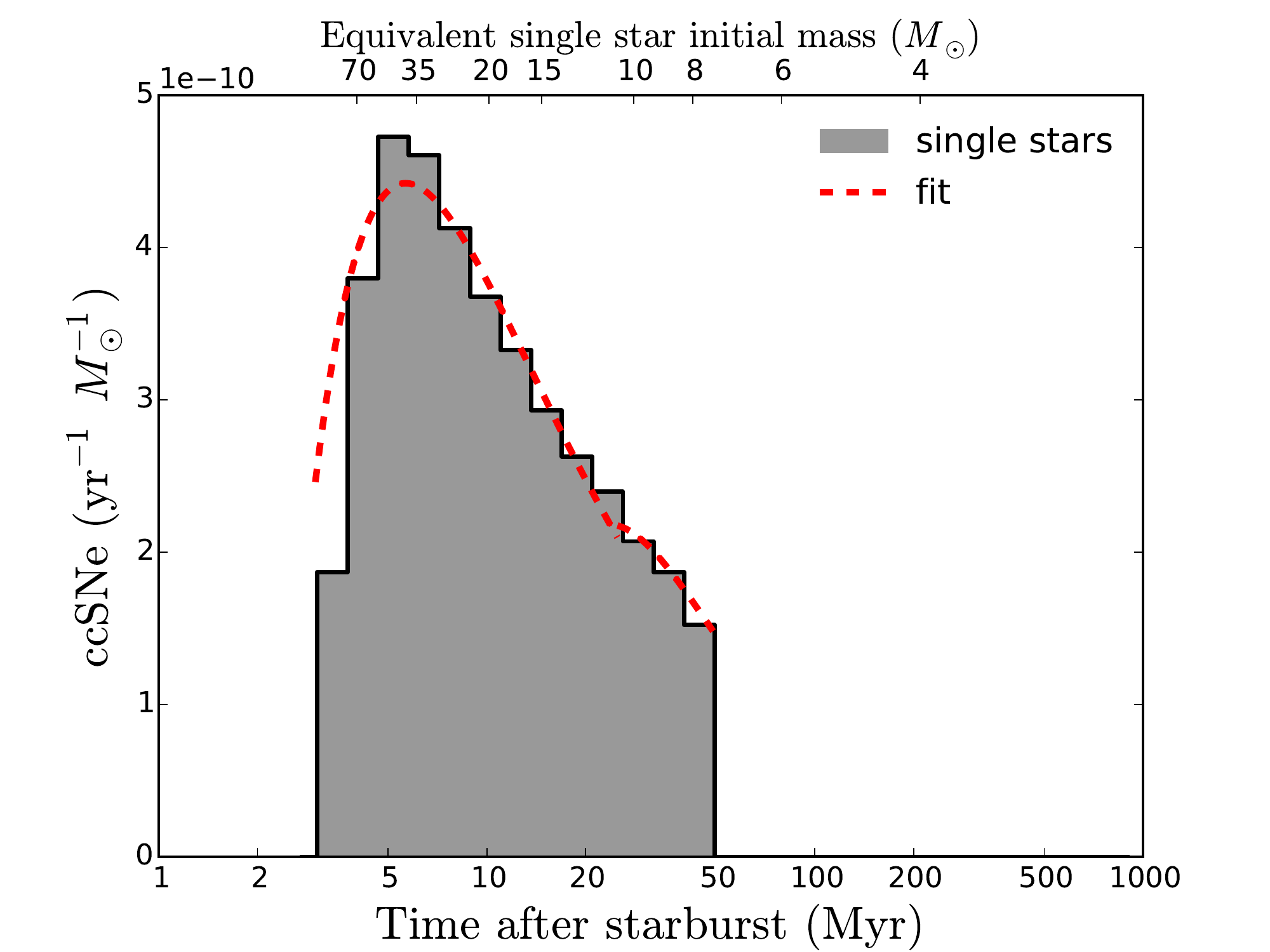}
\includegraphics[width=0.5\textwidth]{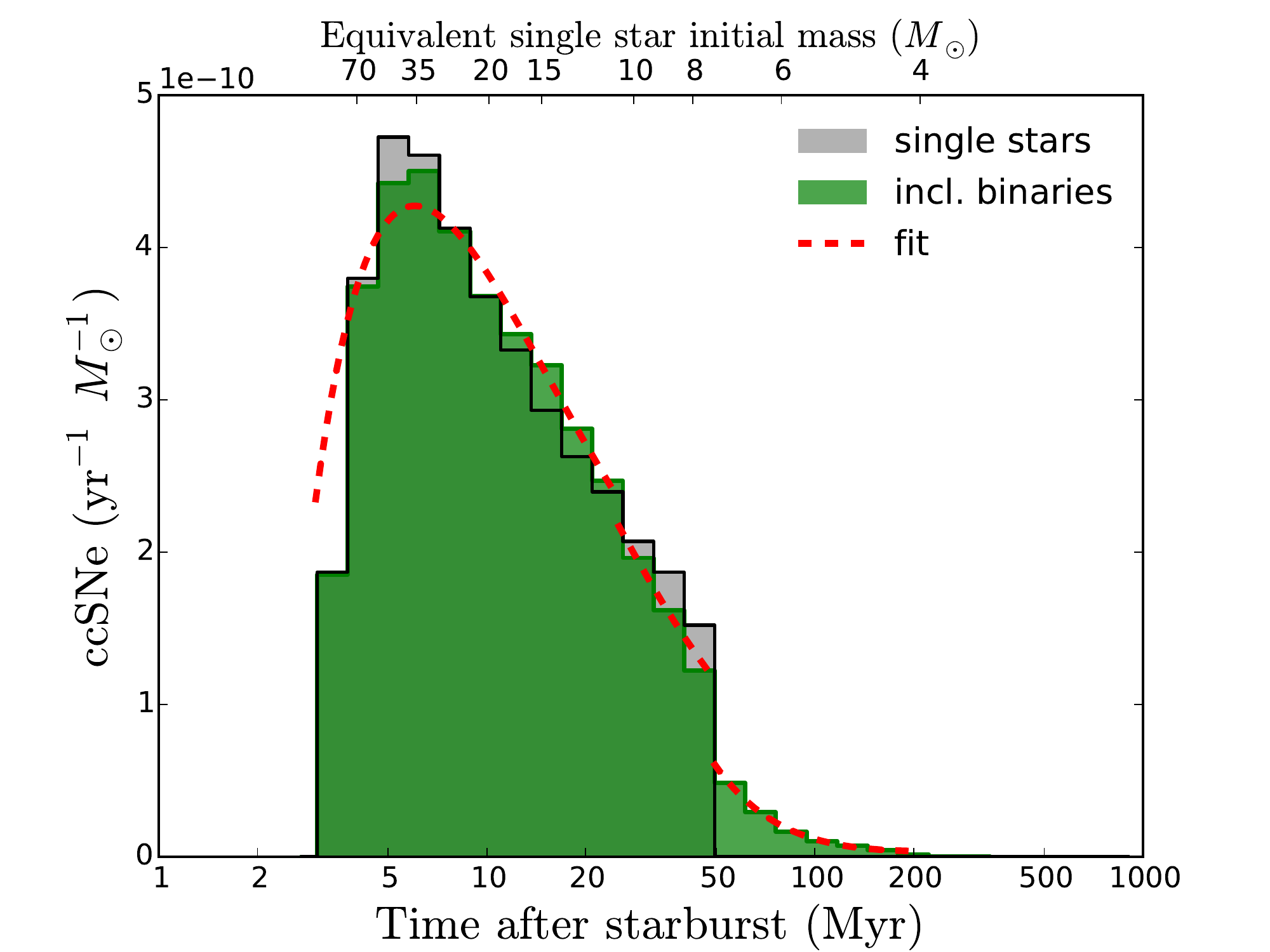}
\caption{ The core-collapse supernova rate as a function of time in a population of single stars (top) and a realistic mix of single stars and binaries (bottom) formed as an instantaneous starburst at $t=0$, for our standard assumptions. The red dashed lines show the fitting formulae. The top axis shows the initial mass of single stars with the corresponding lifetime given in the bottom axis, computed with binary\_c. \label{fig:SN_delay_time_snrate}} 
\end{figure}

In the figures in our results section (Figures \ref{fig:SN_delay_time_single} and \ref{fig:SN_delay_time}), we show the DTD as the number of ccSNe per logarithmic time bin for a $10^6\Msun$ population. These units have the visual advantage that an equal number of systems occupy the same area in different parts of the diagram. However, for various applications of these results, the supernova rate expressed as events per $\mathrm{year}$ per $\Msun$ are more appropriate.
In \autoref{fig:SN_delay_time_snrate}, we provide the supernova rate in our standard simulations containing only single stars (top) and our standard simulation including a realistic fraction of $f_{\mathrm{bin}} = 70\%$ binary systems (bottom).  

For convenience of implementation of our results, we provide fitting functions to our numerical results. In order to derive them, we fitted the cumulative distribution of our supernovae events and then differentiated to derive the supernova rate. The main advantage of this method is that it is independent of the binning. 

The delay-time distribution of the supernova rate, $R$, in units of ccSNe $\rm{yr}^{-1} \Msun^{-1}$  for a single star population is well described by 

\begin{align}
R = 
  \begin{cases}
   10^{-9} [-2.83 + 8.70 (\log_{10}t) -2.07(\log_{10}t)^2](1/t)& \\
   10^{-8} [-4.85 + 6.55 (\log_{10}t) -1.92(\log_{10}t)^2](1/t),&  \\
  \end{cases}
\end{align}
for
\begin{equation*}
\left  \{ 
  \begin{array}{rcl}
 3 \leq &t/\mathrm{Myr}&<25 \mathrm{,}\\
  25\leq &t/\mathrm{Myr}&<48 \mathrm{,} \\
 \end{array}\right.
\end{equation*}
respectively, and $0$ otherwise.

\begin{figure}
\centering
\includegraphics[width=0.5\textwidth]{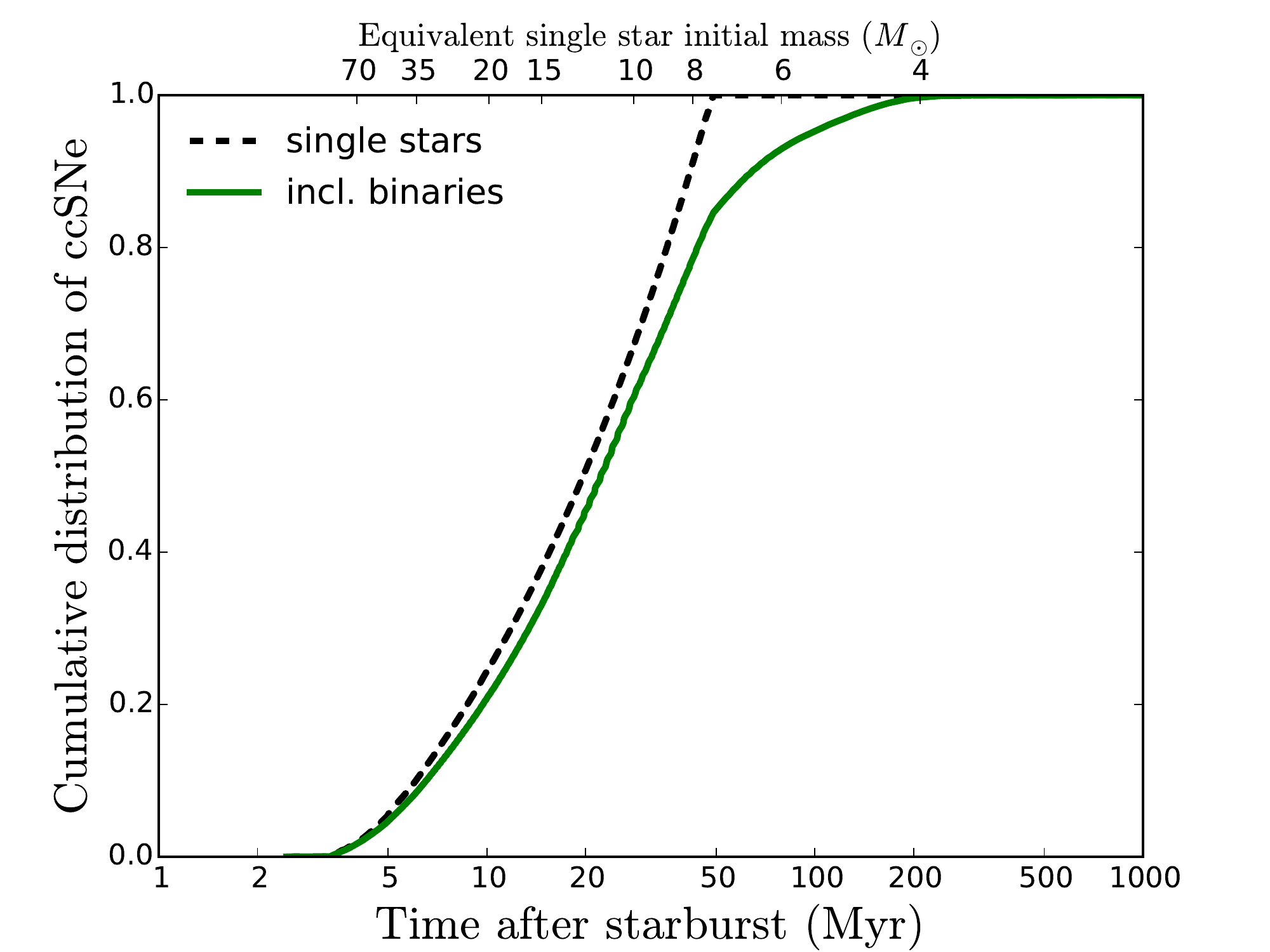}
\caption{ The normalized cumulative distribution of core-collapse supernovae as a function of time in a population of single stars (black dashed line) and a realistic mix of single stars and binaries (green solid line) formed as an instantaneous starburst at $t=0$. The results are shown for our standard assumptions. The top axis shows the initial mass of single stars with the corresponding lifetime given in the bottom axis, computed with binary\_c. \label{fig:cumulative}} 
\end{figure}

The delay-time distribution of the supernova rate, again in units of ccSNe  $\rm{yr}^{-1} \Msun^{-1}$, for a realistic population including $f_{\mathrm{bin}} = 70\%$ binary systems can be approximated as 
\begin{align}
R =
  \begin{cases}
   10^{-9} [-2.65 + 7.51 (\log_{10}t) -0.98(\log_{10}t)^2](1/t)& \\
   10^{-8} [-0.89 + 1.73 (\log_{10}t) -0.51(\log_{10}t)^2](1/t)&  \\
   10^{-8} [\hphantom{-}3.46 - 2.98 (\log_{10}t) +0.65(\log_{10}t)^2](1/t),&  \\
  \end{cases}
\end{align}
for
\begin{equation*}
\left  \{ 
  \begin{array}{rcl}
 3\leq &t/\mathrm{Myr}&<25 \mathrm{,}\\
  25\leq &t/\mathrm{Myr}&<48 \mathrm{,} \\
    48\leq &t/\mathrm{Myr}&<200 \mathrm{,} \\
 \end{array}\right.
\end{equation*}
respectively, and $0$ otherwise.
For the calculation of the fit, we omit ccSNe after $200 \, \mathrm{Myrs}$ because their contribution is not important to the total number of ccSNe (as shown in \autoref{fig:cumulative}).
The relative difference between our fit and our histogram values is, on average, approximately $10\%$.

In \autoref{fig:cumulative}, we show the normalized cumulative distribution both for a single star population and for one including binaries, following our standard assumptions. It can be used to calculate the fraction of supernovae that occurred before a specific time compared to the total.

\section{Properties of the merger channels\label{Appendix:mergers}}

\begin{figure*}[h]
\centering
\includegraphics[width=0.4\textwidth]{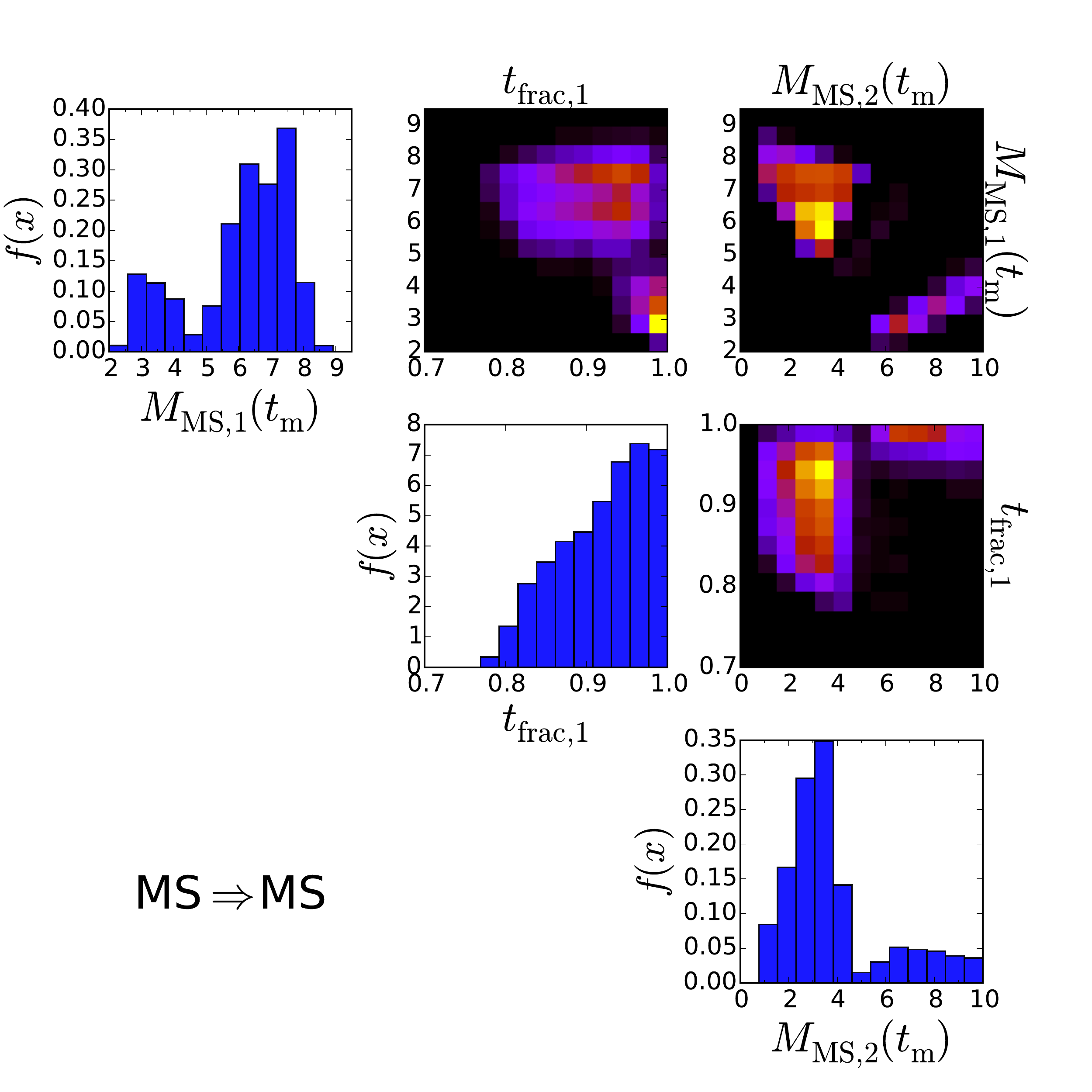} \includegraphics[width=0.4\textwidth]{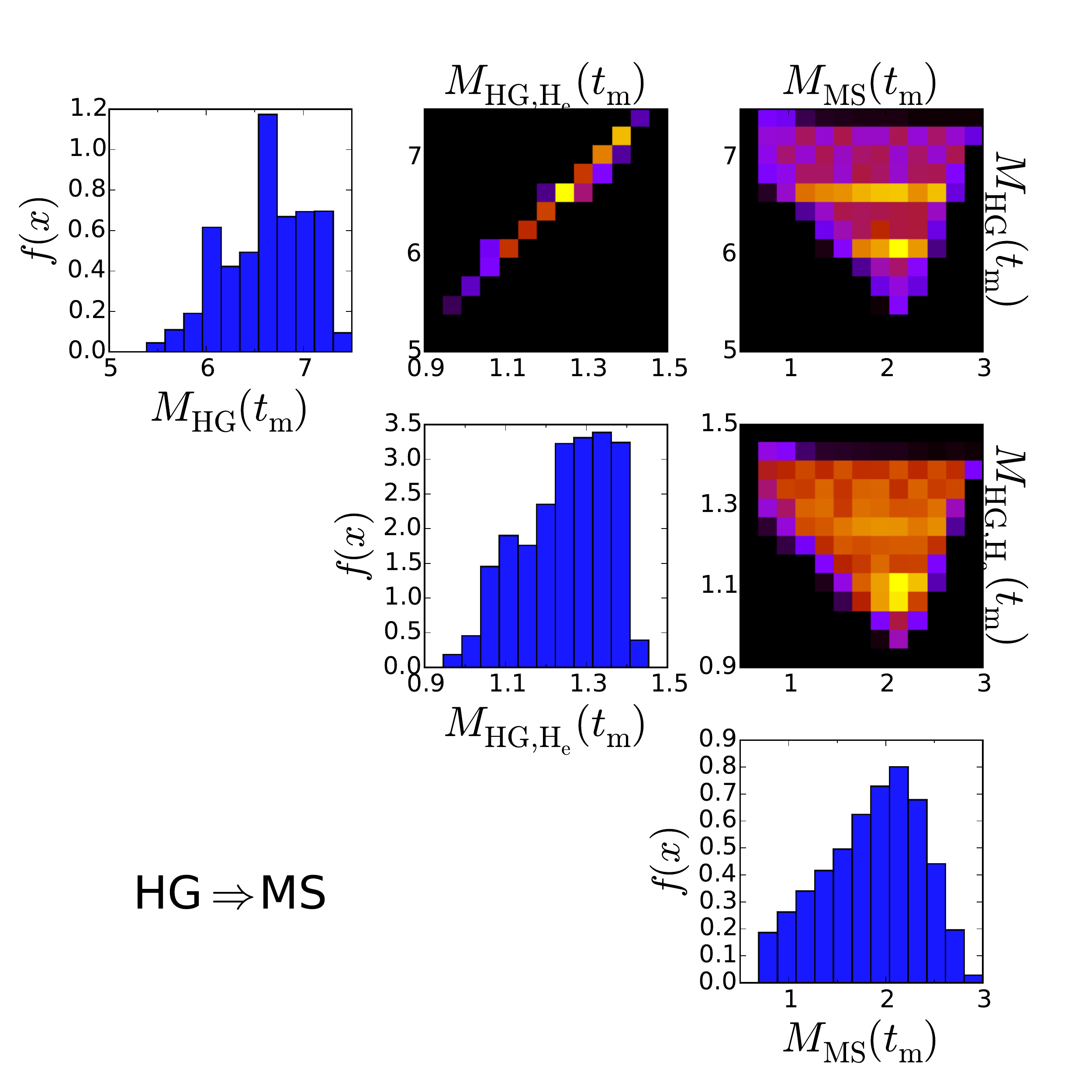}
\includegraphics[width=0.4\textwidth]{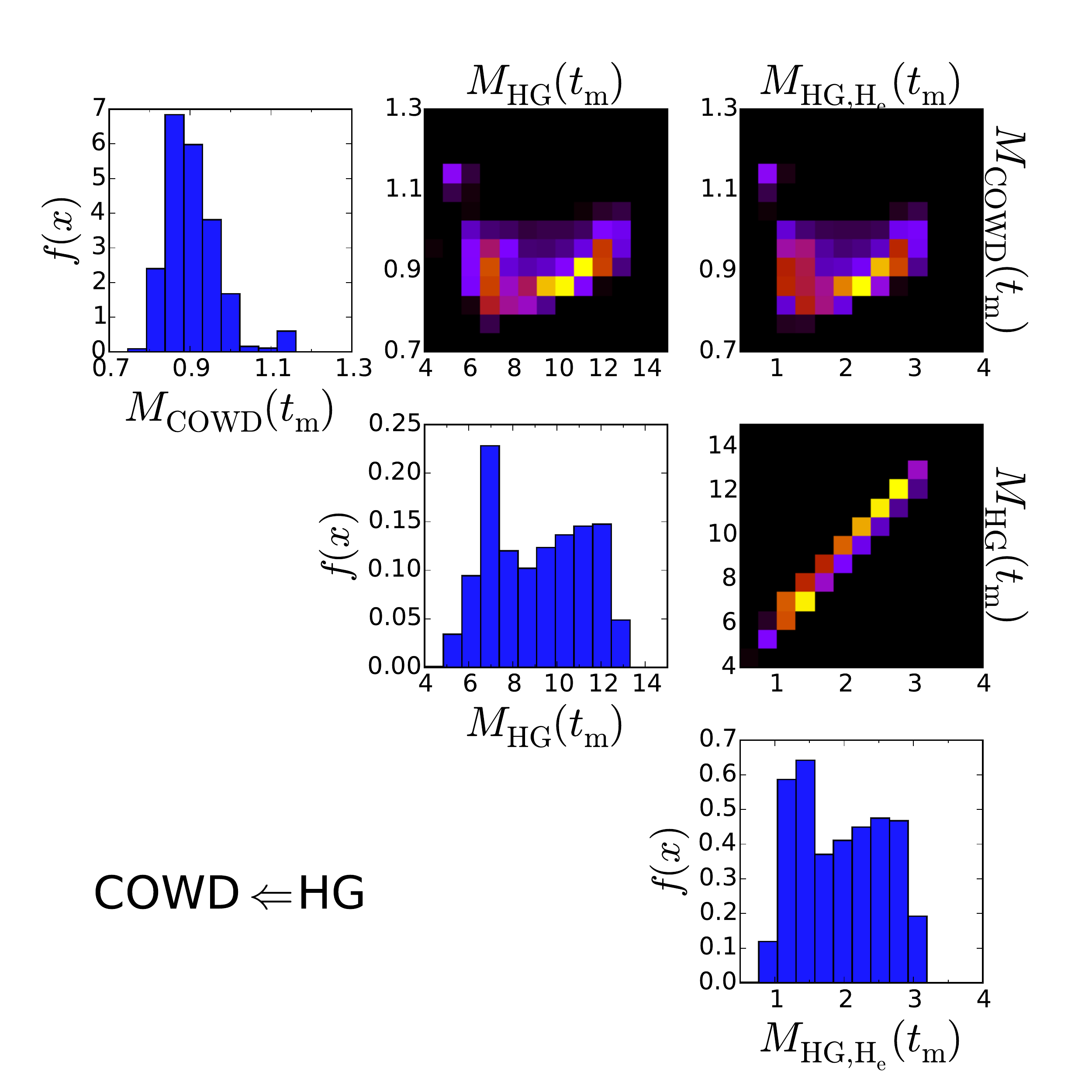}\includegraphics[width=0.4\textwidth]{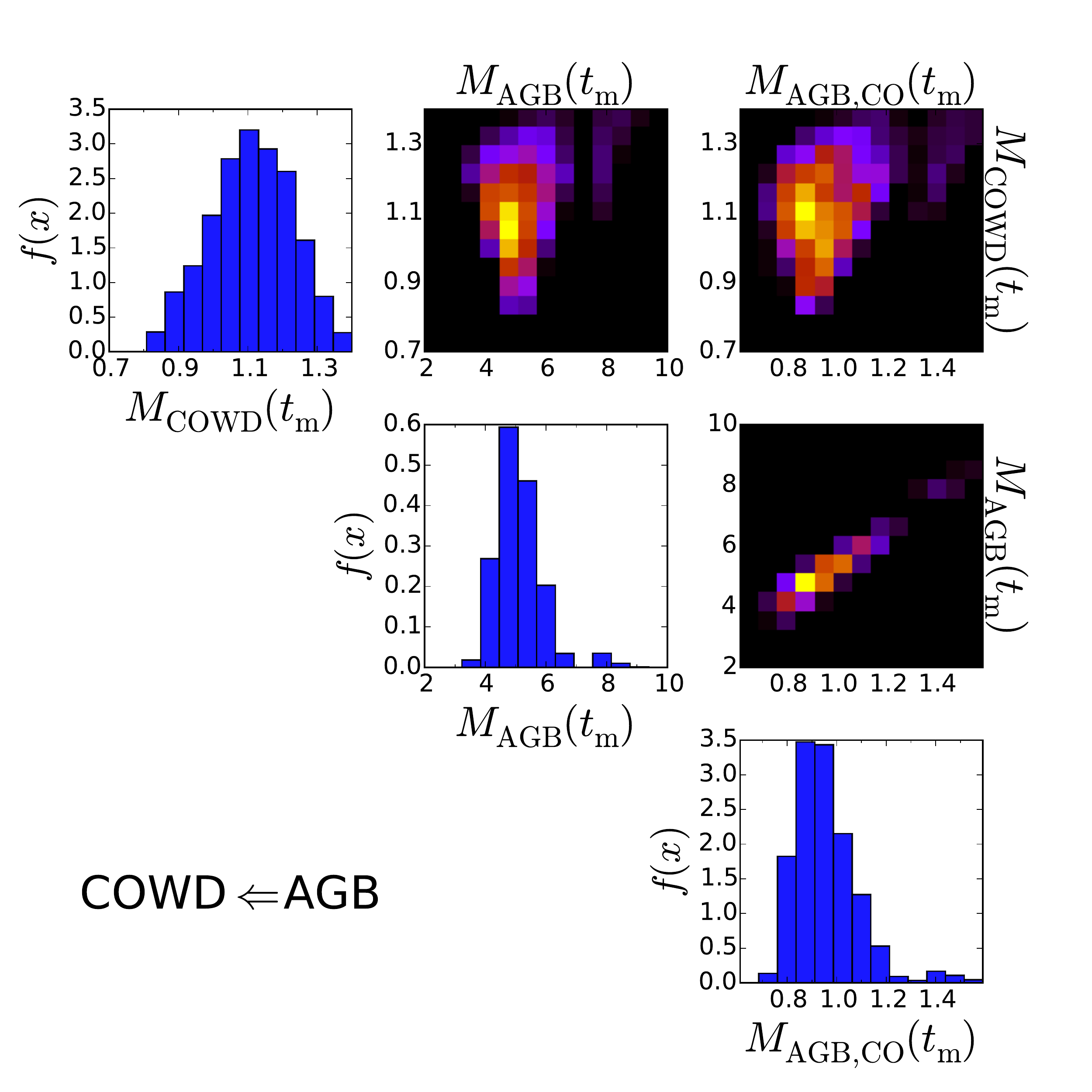}
\includegraphics[width=0.4\textwidth]{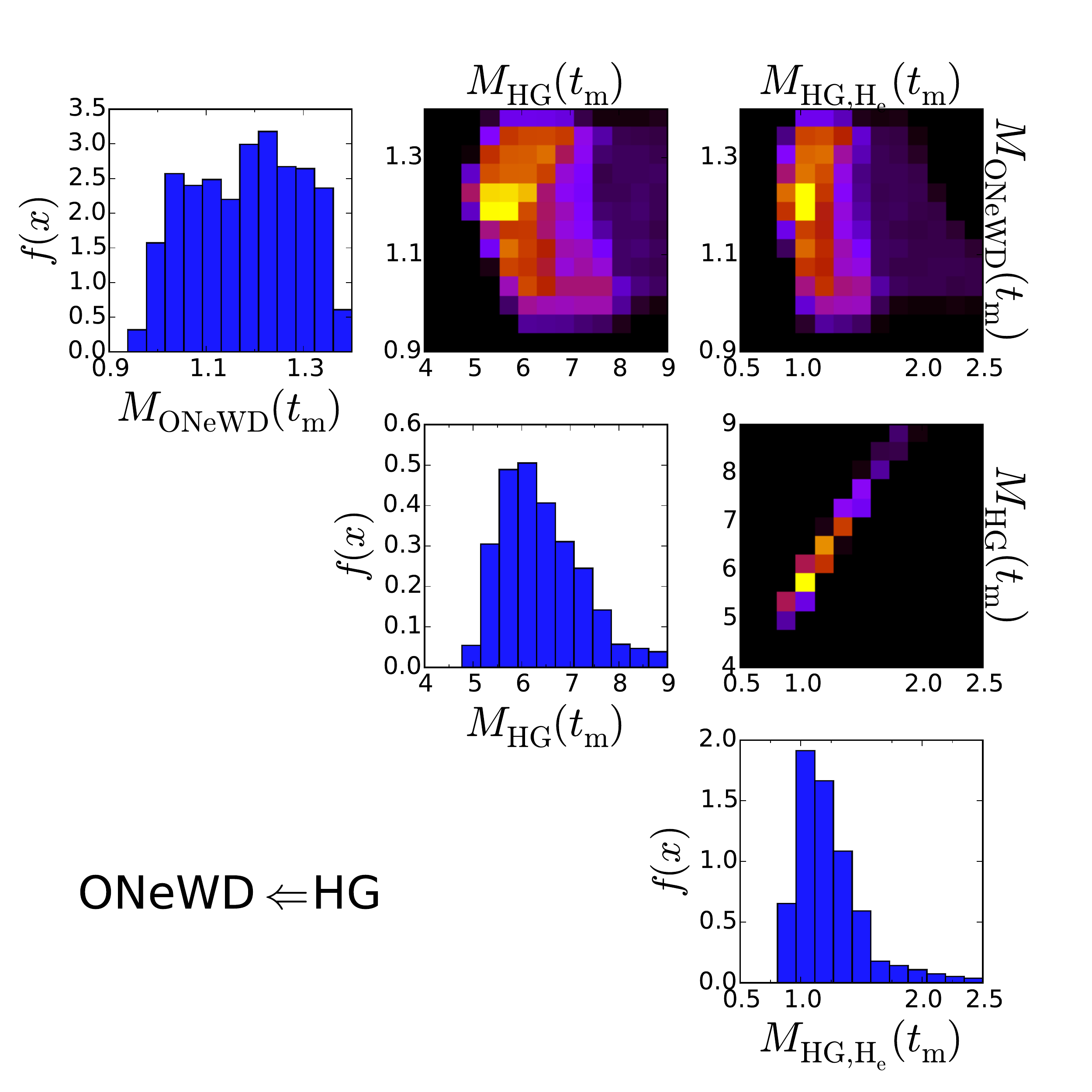}\includegraphics[width=0.4\textwidth]{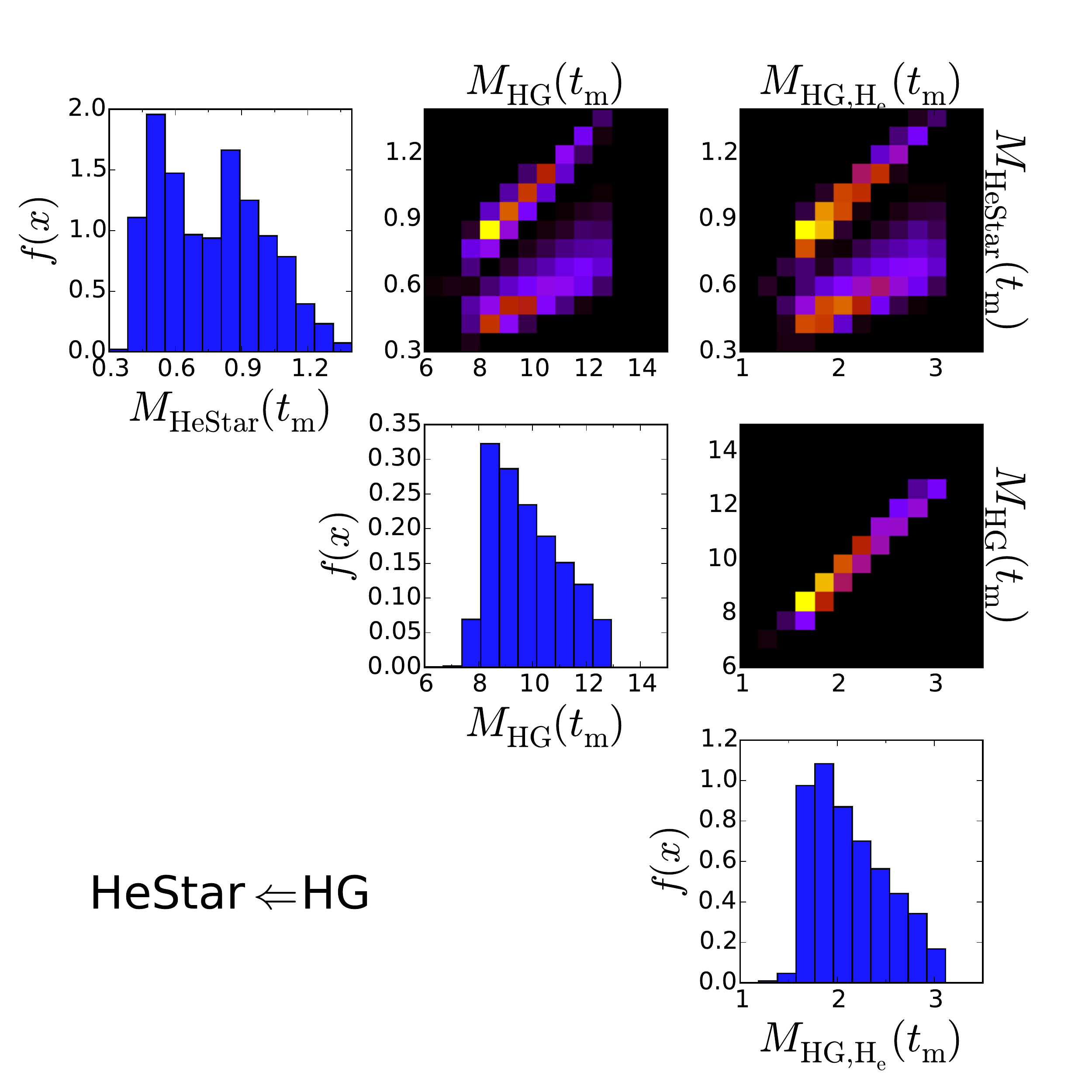}
\caption{ \label{fig:mergers_appendix} These triangular diagrams show the 1D normalized histograms of the properties of the six main subclasses of late events at the moment of merger ($t_m$) and how they are correlated in the 2D histograms. In \autoref{Appendix:mergers}, we describe the various quantities that are shown. Masses are in units of \Msun and $t_{\rm frac}$ is dimensionless.} \end{figure*}

In \autoref{fig:mergers_appendix}, we show six panels of triangular diagrams, one for each of the main merger channels leading to late core-collapse events. These diagrams provide information on the masses and core masses (or evolutionary state) of the two stars involved in the merger. 
The top two panels show the forward mergers, where the merger occurs during a mass transfer event initiated by the primary star. 
These include either two main-sequence stars (top left) or one HG star and a main-sequence star (top right).  The four bottom panels show the properties of various reverse merger channels including those where the primary star is a CO white dwarf (central panels), an ONeMg white dwarf (bottom left), and a helium star (bottom right). In the reverse mergers, the secondary star is typically a HG star. The exception is the central right panel where the secondary star is an AGB star.

The triangular plots in each panel show the 1D distributions of relevant properties together with the corresponding 2D distributions showing how these properties are correlated. The color shading is linear with yellow indicating the highest values and black the lowest.  In each panel, we show the mass of the initially more massive primary at the onset of merging at time $t_m$ in the top row. The mass of the initially less massive secondary is shown on the bottom row.  

The quantity in the central row changes between plots. In all cases that involve a HG star, we show the helium core, $M_{\rm HG,He}$, at the onset of merging.  In the case where  the secondary is an AGB star (central right panel), we show the  CO core mass instead, $M_{\rm AGB,CO}$.  The core mass is not well defined for main-sequence stars. Therefore, we show the fractional age, $t_{\rm frac}$, of the most evolved primary in mergers involving two main-sequence stars (top left panel). We define the fractional age as the age relative to the expected main sequence lifetime of a single star of the same mass. This parameter, which can take values between 0 and 1, is thus an indication of how evolved along the main sequence the star is. 

The channels and implications are discussed in the main text. These diagrams serve as a reference and can in principle be used to guide future detailed simulations of merger products. 

\end{appendix}

\end{document}